\documentclass[%
 reprint,
footinbib,
 amsmath,amssymb,
 aps,
pra,
onecolumn,
notitlepage,
]{revtex4-2}

\usepackage{graphicx}
\usepackage{dcolumn}
\usepackage{bm}
\usepackage{hyperref}
\usepackage[twoside]{fancyhdr} 
\usepackage{lipsum} 
\usepackage{braket} 
\usepackage{amsmath}
\usepackage{amssymb}
\usepackage{dsfont}
\usepackage{nicefrac} 
\usepackage{anyfontsize}

\usepackage{times} 
\usepackage{newtxtext,newtxmath} 

\usepackage{comment} 

\begin{document}

\preprint{APS/123-QED}

\title{Estimating the ground-state hyperfine shifts of group-1 atoms due to long-range collisions, static electric fields, and nearby surfaces}

\author{B. H. McGuyer}
\affiliation{National Coalition of Independent Scholars, Glastonbury, CT 06033, USA}
\author{K. Choksi}
\affiliation{Department of Physics, Lewis \& Clark College, Portland, OR 97219, USA}
\author{N. VonHeeder}
\affiliation{Department of Physics, Lewis \& Clark College, Portland, OR 97219, USA}
\author{B. A. Olsen}
 \email{bolsen@lclark.edu}
 \homepage{https://olsenlab.science}
 \affiliation{Department of Physics, Lewis \& Clark College, Portland, OR 97219, USA}

\date{\today}

\begin{abstract}
Many interactions perturb (or shift) the hyperfine coupling in group-1 atoms between their nuclear spin and single valence electron. 
Previous work provided an approach to estimate this shift for non-reactive collisions with other atoms or molecules at long range, and extended it to estimate the shift from static electric fields. 
We further extend this approach to estimate the shift due to a nearby conductive surface, 
and improve all these estimates by numerically computing updated values for the characteristic energies they use. 
We investigate how to estimate these energies by scaling arguments, and suggest how to adapt the approach  to other applications. 
The results are relevant to pressure shifts in vapor cells, Stark shifts, and wall shifts in nanoscale devices. 

\end{abstract}


\pagestyle{fancy}
\fancyhead{} 
\fancyhead[CO]{\uppercase{Estimating the Ground-state Hyperfine shifts\ldots}}
\fancyhead[CE]{MCGUYER, CHOKSI, VONHEEDER, AND OLSEN}
\fancyfoot{} 
\fancyfoot[CO,CE]{0000XX-\thepage}

{\center \large{APS OPEN SCIENCE {\bf 1}, 0000XX (2026)}\\[4pt] \hrule} 
\vspace{2ex} 

\maketitle

\section{\label{sec:intro}Introduction}

In group-1 atoms (hydrogen and the alkali metals), the hyperfine coupling $A_g {\mathbf I} \cdot {\mathbf S}$ between the spin ${\mathbf I}$ of the nucleus and the spin ${\mathbf S}$ of the single valence electron leads to energy splittings of the ground state corresponding to microwave frequencies. 
These hyperfine frequencies form the basis for many atomic frequency standards (or clocks) and for the current SI definition of the second \cite{bishtPrecisionTimekeepingAtomic2026,vanierQuantumPhysicsAtomic1989}. 
Transitions between ground-state sublevels in alkali-metal atoms split by these frequencies find wide use in quantum metrology \cite{pezzeQuantumMetrologyNonclassical2018} and searches for new physics \cite{safronovaSearchNewPhysics2018,congSpindependentExoticInteractions2025,quintStringentConstraintsNew2026,karshenboimPossibilitiesLaboratorySearches}.
These and other hyperfine interactions allow laser spectroscopy to probe nuclear physics \cite{yangLaserSpectroscopyStudy2023}, enable sensing with nitrogen-vacancy (NV) centers in diamond \cite{duSinglemoleculeScaleMagnetic2024,barrySensitivityOptimizationNVdiamond2020}, and play a critical role in silicon quantum electronics \cite{zwanenburgSiliconQuantumElectronics2013, hansonSpinsFewelectronQuantum2007}. 

Shifts of these group-1 hyperfine frequencies are important to the accuracy, performance, and noise floor of many applications. 
For example, a common technique to improve clock signals is to add non-reactive buffer gasses (often noble gasses) to introduce collisions with alkali-metal atoms \cite{ludlowOpticalAtomicClocks2015}. 
However, these collisions lead to a ``pressure'' shift and broadening of the clock transition, and are the main reason why Rb clocks are secondary standards \cite{margenauPressureEffectsSpectral1936,hindmarshCollisionBroadeningSpectral1973,vanierQuantumPhysicsAtomic1989,vanierQuantumPhysicsAtomic2025,camparoSemiempiricalTheoryCarver2007}. 
Accordingly, pressure shifts have received significant attention over the past several decades. 

\begin{figure}[tb]
    \centering
    \includegraphics[width=510pt]{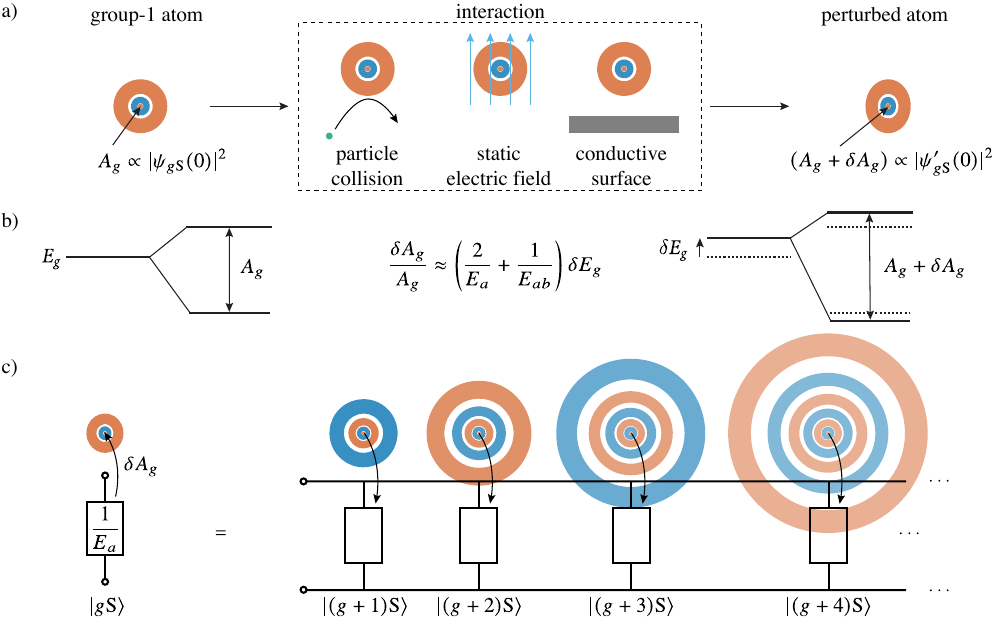}
    \caption{Conceptual picture of hyperfine shifts for a group-1 atom following the form of Eq.~\ref{eq:HFS}. 
(a) When a group-1 atom in its ground state is perturbed by an interaction, such as those illustrated here, the valence electronic wave function changes shape (conceptually sketched for Na). 
If the probability for the valence electron to be at the nucleus changes, then the hyperfine coupling coefficient $A_g$ shifts by a corresponding amount $\delta A_g$. 
(b) For shifts following the form of Eq.~\ref{eq:HFS}, which is repeated here, this shift depends on the overall perturbation $\delta E_g$ as well as the two energies $E_a$ and $E_{ab}$. 
(c) The first energy, $E_a$, only depends on properties of the group-1 atom, so is characteristic of that atom. 
The value of $E_a$ comes from a sum over all $n$S states (see Eq.~\ref{eq:Ea}), with contributions from each excited S-state coupling $A_n$ adding in parallel.} 
    \label{fig:schematic}
\end{figure}

For the ground state, the shift follows the change in the value of the wave function for the group-1 valence electron evaluated at the nucleus, or $\psi_{g\text{S}}(0)$, as illustrated in Fig.~\ref{fig:schematic}. 
Previous work studying collisions and the pressure shift found a simple approach to estimate the shift $\delta A_g$ of the coupling $A_g$ at large collision distances \cite{adrianMatrixEffectsElectron1960,hermanFrequencyShiftsHyperfine1961,vanierQuantumPhysicsAtomic1989}, which has the form 
\begin{align}
\frac{\delta A_g}{A_g}\approx \left(\frac{2}{E_a}+\frac{1}{E_{ab}}\right)\delta E_g\,. 
\label{eq:HFS}
\end{align} 
Here, 
$E_a$ is a ``characteristic'' energy that depends only on the group-1 atom, 
$E_{ab}$ is a smaller correction that depends on the colliding pair, 
and $\delta E_g$ is the collisional perturbation of the ground-state energy $E_g$. 
This form is valid only at large internuclear separations $R$ between the colliding pair, where retardation is negligible. 
More generally, both $\delta A_g$ and $\delta E_g$ are functions of $R$, and $\delta E_g(R)$ is the interaction potential for the pair. 
The full potential $\delta A_g(R)$ is important to the pressure shift 
\cite{ishikawaCollisionalShiftsHyperfine2024,ishikawaFlyingCharacterizationColliding2023,ishikawaNoblegasAtomsCharacterized2022,ishikawaPseudopotentialAnalysisHyperfine2023}, 
as well as alkali-metal--noble-gas van der Waals molecules 
\cite{gongNonlinearPressureShifts2008,mcguyerHyperfineFrequencies872011,mcguyerIsotopeStudyNonlinear2023a} 
and magnetoassociation 
\cite{jones_ultracold_2006,zuchowskiUltracoldRbSrMolecules2010, brueMagneticallyTunableFeshbach2012}.

The original estimates of the energies $E_a$ and $E_{ab}$ in Eq.~\ref{eq:HFS} were fairly crude, such as $E_a\approx \overline{E}_e$ (the mean of certain excited-state energies) \cite{vanierQuantumPhysicsAtomic1989} or $E_a\approx E_\infty$ (the ionization energy) \cite{hermanFrequencyShiftsHyperfine1961}. 
In previous work, we calculated improved values of $E_a$ for Na, K, Rb, and Cs using perturbation theory, and gave an  explicit formula for the accompanying values of $E_{ab}$ \cite{mcguyerHyperfinefrequencyShiftsAlkalimetal2013}. 
Additionally, we extended the approach to estimate hyperfine Stark-shift coefficients $k$ due to static electric fields, and tested our values for $E_a$ by comparing to experimental results. 
Such shifts are important systematics in precision measurements, and many values of $k$ are well known due  to their importance to black-body radiation (BBR) shifts in atomic clocks \cite{itanoShift121982,fleigVariationalCalculationHyperfine2025}. 

In this article, we extend the approach behind Eq.~\ref{eq:HFS} to a new application, the shift from a nearby conductive surface, and extend our calculation of the characteristic energies $E_a$ to all group-1 atoms, by including H, Li, and Fr. 
Using recently expanded experimental results \cite{allegriniSurveyHyperfineStructure2022a}, as well as some improvements to our numerical methods, this provides updated values for Na--Cs. 
As in our previous work, we compute Stark-shift coefficients $k$, now for all group-1 atoms, and compare with experimental results. 
We additionally bolster our results with a scaling argument relating $E_a$ to the splitting $E_S$ between the ground- and first-excited S states of the group-1 atom. 

Our extension to estimate the ground-state hyperfine shifts due to a nearby surface (or wall) treats the van der Waals regime \cite{lennard-jonesProcessesAdsorptionDiffusion1932} of the dispersion forces \cite{klimchitskayaCasimirForceReal2009} between a group-1 atom and the surface. 
This regime applies when the separation $R$ from the surface is large enough to treat the force as that between the atom and the image charges it induces in a conductor, but small enough to ignore retardation \cite{casimirInfluenceRetardationLondonvan1948} and the resulting Casimir-Polder (CP) regime \cite{klimchitskayaCasimirForceReal2009}. 
These surface interactions are receiving increased scrutiny because the size of atomic devices continues to decrease \cite{laliotisAtomsurfacePhysicsReview2021a,wuWallInteractionsSpinpolarized2021}.
In particular, hyperfine shifts in miniature vapor cells \cite{duttaSelectiveReflectionCasimirPolder2025,duttaEffectsHigherorderCasimirPolder2024,duttaProbingCesiumRydberg}, in hollow-core fibers \cite{debordHollowCoreFiberTechnology2019},  or near micromachined waveguides \cite{darosColdAtomsMicromachined2020} can affect spectroscopic measurements and sensor performance.

This article is organized as follows: 
Sec.~\ref{sec:background} outlines how interactions can produce shifts with the form of Eq.~\ref{eq:HFS} through either second- or third-order perturbations. 
We treat three specific applications in detail: wall shifts from nearby conductive surfaces, pressure shifts from long-range collisions, and Stark shifts from static electric fields. 
Sec.~\ref{sec:methods} outlines our approach to estimate the characteristic energy $E_a$ using both numerical calculation and scaling arguments. 
Sec.~\ref{sec:results} reports our results for $E_a$ as well as additional quantities for the three specific applications. 
Sec.~\ref{sec:discussion} discusses these results and possible future work and extensions.  
Sec.~\ref{sec:conclusion} concludes the Article, and 
the Appendix provides reference values for group-1 atoms. 
Our numerical code is available online, along with intermediate calculations and plots \footnote{Code and results are publicly available online here:~\href{https://github.com/olsenlab-science/group1-HFS}{https://github.com/olsenlab-science/group1-HFS}\label{ref:github}}. 

\section{\label{sec:background}Background}

Following Refs.~\onlinecite{mcguyerAtomicPhysicsVaporcell,mcguyerHyperfinefrequencyShiftsAlkalimetal2013}, consider a group-1 atom perturbed by an interaction $W=H_\text{hf}+U$ that is the sum of the contact magnetic-dipole hyperfine interaction of that atom \cite{arimondoExperimentalDeterminationsHyperfine1977a, allegriniSurveyHyperfineStructure2022a}, 
\begin{align}
    H_\text{hf} =\frac{ 8 \pi }{ 3 } g_S \mu_B   \frac{ \mu_I }{ I } \delta({\mathbf r})\, {\mathbf I} \cdot {\mathbf S}\,,
    \label{eq:Hhf}
\end{align} 
and an interaction $U$ with an external perturbation. 
Here, $g_S$ is the g-factor corresponding to the group-1 valence electronic spin ${\bf S}$, $\mu_B$ is the Bohr magneton, $\mu_I$ is the nuclear magneton corresponding to the group-1 nuclear spin ${\bf I}$ with spin quantum number $I$, and $\mathbf r$ is the displacement of the group-1 single valence electron from the nucleus. 
Below, we will use $U$ to model an interaction with either another particle during a collision, an external electric field, or a nearby surface. 
To handle all three cases, let us proceed by assuming a colliding partner, and denoting the state of the colliding pair as $\ket{\mu \nu} = \ket{\mu}_a \otimes \ket{\nu}_b$, with the group-1 atom ``$a$'' in eigenstate $\mu$ and perturbing particle ``$b$'' in state $\nu$, with corresponding energy $E_{\mu\nu}$, and joint ground state $\ket {00}$. 
The cases without a colliding partner follow by setting $\nu = 0$ and using $\Braket{0|0}_b = 1$. 

From time-independent perturbation theory, the first-order energy shift of the ground state due to $W$ is
\begin{align}   \label{EaSum}
    \delta E_{00,1}=\Bra{00} W \Ket{00} = \Braket{00 | H_\text{hf} | 00} + \Braket{00 | U |00}\,
    = \delta E_{00,1}^\text{(hf)} + \delta E_{00,1}^{(U)},
\end{align}
where the first term is the hyperfine energy 
\begin{align} 
\delta E_{00,1}^\text{(hf)} = \Braket{00 | H_\text{hf} | 00} = \Braket{0 | H_\text{hf} |0}_a \otimes \Braket{0|0}_b = A_g\Braket{\mathbf I \cdot \mathbf S}_a \,,
\end{align} 
with a magnetic-dipole hyperfine coupling coefficient $A_g$ for the group-1 atom in the ground (or ``$g$'') S state. 
Here, and subsequently, let us use $g$ to denote the value of the principal quantum number $n$ for the ground state. 
The value of $A_g$ is proportional to the probability density for the valence electron to be at the nucleus: $A_g \propto |\psi_{g\text{S}}(0)|^2$. 
In what follows, we will denote ground-state expectations of operators $X$ as $\Braket{X}_a$ or $\Braket{X}_b$, and occasionally
omit the subscripts ``$a$'' and ``$b$'' for clarity. 

Sec.~\ref{ssec:bkgd_2} treats the case when the second term $\delta E_{00,1}^{(U)} = \Braket{00| U |00} \neq 0$, which leads to a  second-order hyperfine perturbation, 
and Sec.~\ref{ssec:bkgd_3} treats the case of $\delta E_{00,1}^{(U)} = 0$, which leads to a third-order hyperfine perturbation. 
In both cases, we show that the resulting shifts involve a characteristic energy $E_a$ given by:
\begin{align}
    \frac{1}{E_a}=\sum_{n > g} 
		\frac{ \Braket{ g\text{S} | H_\text{hf} | n\text{S} } \Braket{ n\text{S} | r^2 | g\text{S} } }{ \Braket{  H_\text{hf}  }\left(E_{g}-E_{n}\right)\Braket{ r^2}}
        \approx 
        \sum_{n > g} 
        \left( \frac{\sqrt{|A_n/A_g|}}{E_{g}-E_{n }} \right) 
		\frac{  \Braket{ n\text{S} | r^2 | g\text{S} } }{\Braket{ g\text{S} | r^2 | g\text{S} }} 
        > 0\,. 
        \label{eq:Ea}
\end{align} 
Here, the sum is over all excited $n$S states of the group-1 atom with corresponding energies $E_n$, ground state $g$S, and ground-state energy $E_g$.
The second form follows from the square-root formula $\Braket{g\text{S} | H_\text{hf} | n\text{S}} \approx \sqrt{ \Braket{n\text{S} | H_\text{hf} | n\text{S}} \Braket{ H_\text{hf}}} $, which group-1 atoms are expected to satisfy within a percent \cite{bouchiatLinearStarkShifts2008,dzubaOffdiagonalHyperfineInteraction2000}. 
This characteristic energy, which was originally derived in the context of collision shifts \cite{mcguyerAtomicPhysicsVaporcell,mcguyerHyperfinefrequencyShiftsAlkalimetal2013,vanierQuantumPhysicsAtomic1989}, depends only on the properties of the group-1 atom.
Notably, this definition is independent of the perturbing potential $U$. 

\subsection{\label{ssec:bkgd_2}Second-Order Hyperfine Perturbations} 

For the case of  $\delta E_{00,1}^{(U)} \neq 0$, 
and generalizing the approach in Refs.~\onlinecite{mcguyerAtomicPhysicsVaporcell,mcguyerHyperfinefrequencyShiftsAlkalimetal2013}, 
consider the second-order perturbation to the colliding pair ground state energy,
\begin{align}
    \delta E_{00,2} = \sum_{\mu\nu \neq 00}\frac{\left|\Braket{00|W|\mu\nu}\right|^2}{E_{00}-E_{\mu\nu}}\,,
    \label{eq:E_002}
\end{align}
which is a sum over all states but the joint ground state $\Ket{\mu\nu}=\Ket{00}$.
For small $U$, the leading-order perturbation comes from terms linear in both $H_\text{hf}$ and $U$ (note the combinatorial factor of 2): 
\begin{align}
    \delta E_{00,2}^\text{(lin)}  
	&= 	2 \sum_{\mu \nu \neq 00} 
		\frac{ \Braket{ 00 | H_\text{hf} | \mu \nu } \Braket{ \mu \nu | U | 00} }{E_{00}-E_{\mu\nu}} 
	= 	2 \sum_{\mu > 0} 
		\frac{ \Braket{ 00| H_\text{hf} | \mu 0 } \Braket{ \mu 0 | U | 00 } }{E_{00}-E_{\mu 0}}\,. 
        \label{eq:E002lin}
\end{align}
The second form follows from $H_\text{hf}$ acting only on the group-1 atom. 
Note that the only nonzero contributions are from group-1 $n$S states; 
$\Braket{00| H_\text{hf}|\mu 0} \propto \Braket{0| \delta(\mathbf r)|\mu}_a=0$ for all other states.
Thus, in what follows, we will use the integers $g,n$, and corresponding sums, to denote the ground $g$S and excited $n$S states for the group-1 atom.
Next, consider interactions that satisfy (or are well approximated by) the relation 
\begin{align}
    \frac{\Braket{ \mu 0 | U | 00 }}{\Braket{ 00 | U | 00 }} 
    = 
    \frac{\Braket{ \mu 0 | r_a^2 | 00 }}{\Braket{ 00 | r_a^2 | 00 }} 
	\approx 
	\frac{\Braket{ \mu 0 | r^2 | 00 }}{\Braket{ 00 | r^2 | 00 }} 
    \label{eq:scaling}
\end{align} 
for $n$S states, 
where 
$r_a = |{\bf r}_a|$, 
${\bf r}_a = {\bf r} + {\bf r}_c$ is the sum of all group-1 electron positions, 
${\bf r}_c = \sum_i {\bf r}_{c,i}$ is the sum of all group-1 core electron positions, 
and $r = |{\bf r}|$, 
all relative to the group-1 nucleus. 
Then we have 
\begin{align}
    \delta E_{00,2}^\text{(lin)}  
	&\approx 	
        2 \Braket{ 00 | U | 00 } \sum_{n > g} 
		\frac{ \Braket{ g| H_\text{hf} | n } \Braket{ n | r^2 | g} }{\left(E_{g0}-E_{n 0}\right)\Braket{ r^2}}  
    & = 2 \delta E_{00,1}^{(U)} \Braket{  H_\text{hf}  }
          \sum_{n > g} 
		\frac{ \Braket{ g| H_\text{hf} | n } \Braket{ n | r^2 | g} }{ \Braket{  H_\text{hf}  }\left(E_{g0}-E_{n 0}\right)\Braket{ r^2}}
        = \delta E_{00,1}^{(U)} \delta E_{00,1}^\text{(hf)} 
         \left( \frac{2}{E_a} \right), 
\end{align}
which recovers the characteristic energy $E_a$ of Eq.~\ref{eq:Ea}. 
This allows us to write the fractional ground-state hyperfine shift in the form of Eq.~\ref{eq:HFS} as 
\begin{align}  
    \frac{\delta A_g}{A_g}
    \approx \frac{\delta E_{00,2}^\text{(lin)} }{\delta E_{00,1}^\text{(hf)}} 
    = \left( \frac{2}{E_a} \right) \delta E_{00,1}^{(U)}\,.
    \label{eq:Ea2ndOrderResult}
\end{align}
The only requirements for this result are (i) that $\Braket{00|U|00}\neq0$ and (ii) that the relationship in Eq.~\ref{eq:scaling} holds. 

Before we continue, note that, somewhat surprisingly, this result does not include the $E_{ab}$ term present in Eq.~\ref{eq:HFS}. 
As a result, if $U$ models a collision, for example, then any relevant properties of the collision partner would only contribute via $\delta E_{00,1}^{(U)}$. 
Our treatment of collisions below does not satisfy the requirements for this form, since $\Braket{00|U|00} = 0$ for that application. 
However, as noted in Ref.~\cite{mcguyerHyperfinefrequencyShiftsAlkalimetal2013}, an effective operator can be constructed that does satisfy the requirements for this form, as shown in Sec.~\ref{ssec:bkgd_3}, which can be used to model collisions or other interactions. 
Therefore, care is needed using such effective operators because they neglect the $E_{ab}$ term. 

\subsubsection{\label{sssec:bkgd_3surf}Wall shift from a nearby surface}

\begin{figure}[bt]
    \centering
    \includegraphics[width=17cm]{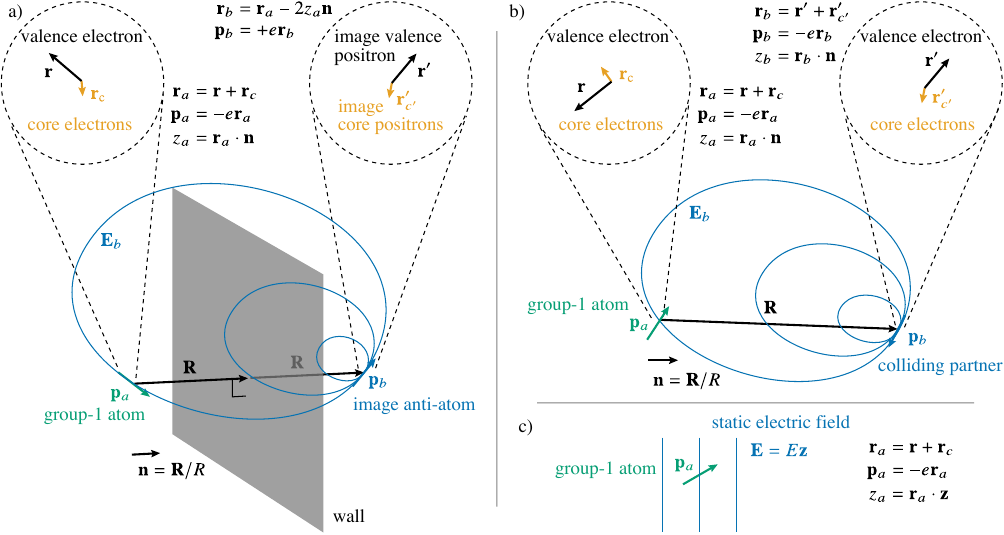}
    \caption{Diagrams for applications in Section~\ref{sec:background}. 
    (a) Wall shift from a nearby conductive surface. The group-1 atom at a distance $\mathbf{R}$ from the wall  induces an image anti-atom $2\mathbf{R}$ away. The  image anti-atom has instantaneous electric dipole moment $\mathbf{p}_b$, producing an electric field $\mathbf{E}_b$ which interacts with the atom's  moment $\mathbf{p}_a$.
    (b) Pressure shift from a long-range collision. The partner's instantaneous electric dipole moment $\mathbf{p}_b$ produces an electric field $\mathbf{E}_b$ which interacts with the group-1 atom's moment $\mathbf{p}_a$.
    (c) Stark shift from a static electric field. We define coordinates so $\mathbf{E} = E \mathbf{z}$.
    All notation is defined in the text. 
    \label{fig:setups}}
\end{figure}

One example of a shift with the form of Eq.~\ref{eq:Ea2ndOrderResult} is the interaction of  a group-1 atom with a nearby conducting wall (c.f.\ Complement $\text{C}_\text{XI}$ Part 4 of Ref.~\onlinecite{cohen-tannoudjiQuantumMechanics22005}, as depicted in Fig.~\ref{fig:setups}a). 
Consider a group-1 atom at position $\mathbf {x}_a$, which is a normal distance $\mathbf R$ away from a flat, perfectly conducting surface, with $\mathbf n = \mathbf R / |\mathbf R|$ the unit vector pointing from the atom towards the closest point on the surface. 
The atom induces surface charges that produce an electric field equivalent to an image anti-atom at position $\mathbf{x}_b = \mathbf{x}_a + 2 \mathbf{R}$.
Assuming the distance $R = |\mathbf R|$ is large enough to neglect details of the surface, but small enough to neglect retardation effects \cite{casimirInfluenceRetardationLondonvan1948}, the leading-order interaction in $U$ has a dispersive van der Waals form given by 
\begin{align}
    U_S=-\frac{1}{2}\mathbf p_a \cdot \mathbf E_b(\mathbf x_a)\,,
\end{align}
where $\mathbf p_a = -e \mathbf r_a$ is the instantaneous electric dipole moment of the atom, 
${\bf E}_b$ is the electric field produced by the image, 
and the coefficient $\nicefrac{1}{2}$ removes the fictitious energy behind the wall. 
The field $\mathbf E_b(\mathbf x_a) = (3 \mathbf n \mathbf n - \mathds{1}) \cdot \mathbf p_b/(2R)^3$, where $\mathds{1}$ is the identity dyadic tensor and $\mathbf p_b = + e \mathbf r_b$ is the instantaneous electric dipole moment of the image anti-atom. 
As sketched in Fig.~\ref{fig:setups}a, the sum of all anti-atom electronic positions $\mathbf r_b = \mathbf r_a-2z_a \mathbf n$, where the normal projection $z_a = \mathbf r_a \cdot \mathbf n$. 
Together, this gives 
\begin{align}
    U_S&= \frac{e^2}{2} \left[\frac{\mathbf r_a  \cdot(3 \mathbf n \mathbf n - \mathds{1}) \cdot (\mathbf r_a - 2 z_a \mathbf n) }{8R^3}\right] 
    = \frac{-e^2}{16R^3}(r_a^2+z_a^2)\,.
\end{align} 
Noting that $\Braket{U_S} \neq 0$ and that $\Braket{n|z_a^2|g} = \nicefrac{1}{3}\Braket{n|r_a^2|g}$ by symmetry for $n$S states, we see that both required conditions are satisfied for the result of Eq.~\ref{eq:Ea2ndOrderResult}. 
Rewriting the first-order perturbation as 
$\delta E_{00,1}^{(U_S)} = \Braket{U_S} =-C_{3}/R^3$, where  
\begin{align} 	\label{eq:C3Trick}
C_{3} = e^2\Braket{r_a^2}/12 
\end{align} 
is the ground-state, atom-surface dispersion coefficient $C_3$ of the group-1 atom \cite{dereviankoElectricDipolePolarizabilities2010a}, 
we can write the shift as 
\begin{align}
    \frac{\delta A_{g,S}}{A_g} 
    \approx 
        - \left( \frac{2}{E_a}\right) \frac{C_{3}}{R^3}\,. 
    \label{eq:HFS_wall} 
\end{align}
We estimate this ``wall'' (or ``surface'') shift in Section~\ref{ssec:surface} for a representative distance $R$. 

\subsection{\label{ssec:bkgd_3}Third-order Hyperfine Perturbations}

For the case of $\delta E_{00,1}^{(U)}= 0$, and generalizing the approach in 
Refs.~\onlinecite{mcguyerAtomicPhysicsVaporcell,mcguyerHyperfinefrequencyShiftsAlkalimetal2013}, 
consider an interaction $U$ that has a nonzero second-order perturbation, 
\begin{align}
        \delta E_{00,2}^{(U)} = \sum_{\mu\nu \neq 00}\frac{\left|\Braket{00|U|\mu\nu}\right|^2}{E_{00}-E_{\mu\nu}}\neq0\,, 
        \label{eq:E002}
\end{align}
and that satisfies $\Braket {00|U|\mu 0}=0$ when $\mu$ is an $n$S state of the group-1 atom, say from spherical symmetry. 
Then the only non-vanishing terms of the total second-order perturbation $\delta E_{00,2}$ are quadratic in either $H_\text{hf}$ or $U$, with no mixed terms linear in both.
Instead, the third-order shift 
\begin{align}
    \delta E_{00,3} 
    &= 	\sum_{\mu \nu \neq 00} \sum_{\eta \xi \neq 00} 
	\frac{ \Braket{ 00| W | \mu \nu } 
	\Braket{ \mu \nu | W | \eta \xi } 
	\Braket{ \eta \xi | W | 00 }}{(E_{00}-E_{\mu \nu})(E_{00}-E_{\eta \xi})} 
	- \Braket{ 00 | W | 00 } 
	\sum_{\rho \sigma \neq 00} 
	\frac{|\Braket{ 00| W | \rho \sigma }|^2}{(E_{00}-E_{\rho \sigma})^2}\,,
\end{align}
contains terms linear in $H_\text{hf}$ (note the combinatorial factor of 2): 
\begin{align}
    \delta E_{00,3}^\text{(lin)} 
    &= 	2\sum_{\mu \nu \neq 00} \sum_{\eta \xi \neq 00} 
	\frac{ \Braket{ 00| H_\text{hf} | \mu \nu } 
	\Braket{ \mu \nu | U | \eta \xi } 
	\Braket{ \eta \xi | U | 00 }}{(E_{00}-E_{\mu \nu})(E_{00}-E_{\eta \xi})} 
	- \Braket{ 00 | H_\text{hf} | 00 } 
	\sum_{\rho \sigma \neq 00} 
	\frac{|\Braket{ 00| U | \rho \sigma }|^2}{(E_{00}-E_{\rho \sigma})^2}\,.
\end{align} 
As before, we can simplify the first sum by noting that $H_\text{hf}$ acts only on the group-1 atom, giving $\nu = 0$. 
To proceed, let us approximate the denominators with an effective energy scale 
$\hat{E}$ 
using the substitutions
$(E_{00}-E_{\mu \nu})(E_{00}-E_{\eta \xi})\approx - \hat{E}(E_{00}-E_{\mu \nu})$ and 
$(E_{00}-E_{\rho \sigma})^2\approx - \hat{E}(E_{00}-E_{\rho \sigma})$. 
Then we can remove the next sum with the closure relation $\sum_{\eta \xi} \ket{\eta \xi} \bra{\eta \xi}=1$, noting $\Braket{ 00| U | 00} = 0$ by assumption. 
Next, consider interactions that satisfy (or are well approximated) by the relation 
\begin{align}
    \frac{\Braket{ \mu 0 | U^2 | 00 }}{\Braket{ 00 | U^2 | 00 }} 
    = 
	\frac{\Braket{ \mu 0 | r_a^2 | 00 }}{\Braket{ 00 | r_a^2 | 00 }} 
	\approx 
	\frac{\Braket{ \mu 0 | r^2 | 00 }}{\Braket{ 00 | r^2 | 00 }} 
    \label{eq:scaling2}
\end{align}
for $n$S states. 
All together, we then have 
\begin{align}
    \delta E_{00,3}^\text{(lin)} 
    &\approx 	- 2\Braket{ 00 | U^2 | 00 } \sum_{\mu \neq 0} 
	\frac{ \Braket{ 00| H_\text{hf} | \mu 0 } 
	 \Braket{ \mu 0 | r^2 | 00 }}
    {\hat{E}(E_{00}-E_{\mu \nu})\Braket{ 00 | r^2 | 00 }} 
	+ \delta E_{00,1}^{(\text{hf})} 
	\sum_{\rho \sigma \neq 00} 
	\frac{\left|\Braket{ 00| U | \rho\sigma }\right|^2}{\hat{E}(E_{00}-E_{\rho \sigma})} \nonumber  \\
    &= 	
    	\left( \frac{1}{\hat{E}} \right) \left( - 2\Braket{ 00 | U^2 | 00 } \Braket{H_\text{hf}} \sum_{\mu \neq 0} 
	\frac{ \Braket{ 00| H_\text{hf} | \mu 0 } 
	 \Braket{ \mu 0 | r^2 | 00 }}
    {\Braket{H_\text{hf}}(E_{00}-E_{\mu \nu})\Braket{ r^2 }} 
	+ \delta E_{00,1}^{(\text{hf})} 
	\delta E_{00,2}^{(U)} \right) \nonumber  \\ 
    &= 	
    \left( \frac{1}{\hat{E}} \right) \left( - \Braket{ 00 | U^2 | 00 }  \frac{2}{E_a}  
	+ 
	\delta E_{00,2}^{(U)} \right) \delta E_{00,1}^{(\text{hf})}  \,, 
\end{align}
which recovers the characteristic energy $E_a$ of Eq.~\ref{eq:Ea}.  
The resulting fractional ground-state hyperfine shift is 
\begin{align}
    \frac{\delta A_g}{A_g}
    \approx \frac{\delta E_{00,3}^\text{(lin)} }{\delta E_{00,1}^\text{(hf)}} 
    \approx \left( \frac{1}{\hat{E}} \right) \left(-\Braket{ 00 | U^2 | 00 } \frac{2}{E_a}+\delta E_{00,2}^{(U)}\right)\,, 
    \label{eq:E002int}
\end{align}
which is similar to but not exactly of the same form as Eq.~\ref{eq:HFS}. 
This intermediate form is used below for calculating the Stark shift. 
Otherwise, lacking any particular motivation for choosing $\hat{E}$, we can choose a self-consistent value for which the approximation 
$E_{00}-E_{\rho \sigma} \approx - \hat{E}$ used above, together with closure, returns the $\delta E_{00,2}^{(U)}$ of Eq.~\ref{eq:E002} to itself. 
This self-consistent value for $\hat{E}$ is 
\begin{align} 	\label{eq:DefineEab}
E_{ab}=-\frac{\Braket{00|U^2|00}}{\delta E_{00,2}^{(U)}}\,. 
\end{align} 
After substituting this value for $\hat{E}$, the intermediate form of Eq.~\ref{eq:E002int} recovers the form of Eq.~\ref{eq:HFS}: 
\begin{align}
    \frac{\delta A_g}{A_g}
     \approx \left(\frac{2}{E_a}+\frac{1}{E_{ab}}\right)\delta E_{00,2}^{(U)}\,.
    \label{eq:HFS_Eab}
\end{align} 
As before, the characteristic energy $E_a$ depends only on properties of the group-1 atom, but here the ``correction'' energy $E_{ab}$  depends on properties of both the atom and the perturbation. 
The only requirements for both this final result as well as the intermediate result of Eq.~\ref{eq:E002int} are  
(i) that $\Braket{00|U|00} = 0$, 
(ii) that $\delta E_{00,2}^{(U)} \neq 0$, 
(iii) that $\Braket{00|U|\mu0}=0$ for $n$S states $\ket{\mu}_a$, 
and (iv) that the relation in Eq.~\ref{eq:scaling2} holds. 

Before we continue, let us return to the comment following Eq.~\ref{eq:Ea2ndOrderResult} about $E_{ab}$ being absent in the second-order result. 
Note that the effective operator  
$U'=\sum_{\mu\nu\neq 00} U \ket{\mu \nu} \bra{\mu \nu} U / (E_{00}-E_{\mu\nu})$, which could come from a Van Vleck, contact, or related transformation, 
reproduces the leading perturbation of $U$ as a first-order perturbation: 
$\delta E_{00,1}^{(U')}=\delta E_{00,2}^{(U)}$. 
Using closure, this operator also satisfies the relation in Eq.~\ref{eq:scaling}. 
Therefore, this operator satisfies the requirements for the second-order result, so only partially recovers the third-order shift of Eq.~\ref{eq:HFS_Eab} as a second-order shift of Eq.~\ref{eq:Ea2ndOrderResult}, which has no $E_{ab}$ term. 

\subsubsection{\label{ssec:pressure} Pressure shifts from long-range collisions}

The primary application that derived $E_a$ was modeling a non-reactive, long-range collision between a group-1 atom and another atom or molecule. 
To proceed, let us consider the typical case of a collision between a group-1 atom and a group-18 (noble gas) atom, which has a closed shell, and neglect the nuclear spin of the group-18 atom. 
Collisions with open-shell atoms or molecules involve spin interactions between the pair, which may be significant. 
Consider a group-1 atom at position $\mathbf x_a$, and a perturbing particle at position $\mathbf x_b = \mathbf x_a + \mathbf R$ (see Fig.~\ref{fig:setups}b), both in their ground state.
For $R=|\mathbf R|$ large compared to either ground-state wave function extent, but small enough to neglect retardation effects \cite{casimirInfluenceRetardationLondonvan1948}, the leading-order interaction in $U$ is given by a dispersive van der Waals form, 
\begin{align}
    U_C = -\mathbf p_a \cdot \mathbf E_b(\mathbf x_a)\,,
\end{align} 
where, as before, $\mathbf p_a =-e \mathbf r_a$. 
Likewise, the field $\mathbf E_b(\mathbf x_a) = (3 \mathbf n \mathbf n - \mathds{1}) \cdot \mathbf p_b /R^3$, but here  $\mathbf p_b =-e \mathbf r_b$ is the instantaneous electric dipole moment of the group-18 atom, and $\mathbf r_b$ the sum of its electron positions relative to its nucleus. 
Together, this gives 
\begin{align}
    U_C &= - e^2 \left[\frac{ \mathbf r_a \cdot (3\mathbf n \mathbf n - \mathds{1}) \cdot \mathbf r_b }{R^3} \right] 
     =\frac{e^2}{R^3} (\mathbf r_a \cdot \mathbf r_b-3z_a z_b)\,, 
\end{align}
where the projections 
$z_a = \mathbf r_a \cdot \mathbf n$ 
and $z_b = \mathbf r_b \cdot \mathbf n$. 
The first-order perturbation $E_{00,1}^{(U_C)}$ vanishes because $\Braket{r_a} = \Braket{z_a} = 0$ from symmetry. 
Similarly, $\Braket{00|U_C | \mu0}=0$ due to spherical symmetry for $n$S states. 
Noting again that $\Braket{n|z_a^2|g} = \nicefrac{1}{3}\Braket{n|r_a^2|g}$ for $n$S states, we see that the interaction satisfies Eq.~\ref{eq:scaling2}. 
Assuming the second-order perturbation $E_{00,2}^{(U_C)} \neq 0$, all requirements are met for the result of Eq.~\ref{eq:HFS_Eab}. 
Rewriting the second-order perturbation as 
$\delta E_{00,2}^{(U_C)} = -C_6 R^{-6}$, where 
$C_6$ 
is the van der Waals dispersion coefficient of the colliding pair \cite{dereviankoElectricDipolePolarizabilities2010a,mitroyTheoryApplicationsAtomic2010a, zhangLongrangeDispersionInteractions2007}, 
we can write the shift as  
\begin{align} 
    \frac{\delta A_{g,C}}{A_g} 
    \approx 
        - \left( \frac{2}{E_a} + \frac{1}{E_{ab}} \right) \frac{C_{6}}{R^6}\,, 
\end{align}
where the correction energy of Eq.~\ref{eq:DefineEab} is 
\begin{align}
    E_{ab} = -\frac{\Braket{ 00| U_C^2 |00}}{\delta E_{00,2}^{(U_C)}} \approx \frac{2 e^4 \Braket{ r_a^2 } \Braket{ r_b^2}}{3 \,C_6},
\end{align}
assuming uncorrelated electron positions \cite{mcguyerHyperfinefrequencyShiftsAlkalimetal2013}. 
Using Eq.~\ref{eq:C3Trick} and approximating $\braket{r_a^2}\approx \braket{r^2}$, we can rewrite this as 
\begin{align}   \label{eq:EabCollision}
    E_{ab} \approx \frac{8e^2 \braket{r^2} C_{3,b}}{C_{6}}, 
\end{align} 
which uses an atom-surface $C_3$ dispersion coefficient for the collision partner. 
We estimate the $E_{ab}$ of Eq.~\ref{eq:EabCollision} in Section~\ref{ssec:Eab} for pairs of group-1 and group-18 (noble gas) atoms. 

\subsubsection{\label{ssec:stark} Stark shift from a static electric field}

To test values of the characteristic energies $E_a$ for pressure shifts, Ref.~\onlinecite{mcguyerHyperfinefrequencyShiftsAlkalimetal2013} connected $E_a$ with scalar, static Stark shifts. 
For this application, consider a group-1 atom in a uniform, static electric field 
$\mathbf E = E \mathbf z$ aligned with the Cartesian unit vector $\mathbf z$ (see Fig.~\ref{fig:setups}c). 
The dominant interaction responsible for this shift has the form 
\begin{align}
    U_E=-\mathbf p_a \cdot \mathbf E = eEz_a\,,
\end{align}
where again $\mathbf p_a = -e \mathbf r_a$ and the projection $z_a = \mathbf r_a \cdot \mathbf z$. 
The first-order perturbation $E_{00,1}^{(U_E)}$ vanishes because $\Braket{z_a} = 0$, 
and $\Braket{00|U_E | \mu0}=0$ for $n$S states, both from symmetry. 
Using $\Braket{n|z_a^2|g} = \nicefrac{1}{3}\Braket{n|r_a^2|g}$ for $n$S states, this interaction satisfies Eq.~\ref{eq:scaling2}. 
Therefore, assuming the second-order perturbation $E_{00,2}^{(U_E)} \neq 0$, all requirements are met for the third-order shift results. 
Rewriting the second-order perturbation as 
$\delta E_{00,2}^{(U_E)} = -\alpha_a(0)E^2/2$, where 
$\alpha_a(0)$ is the ground-state static polarizability of the group-1 atom \cite{vanierQuantumPhysicsAtomic1989,mitroyTheoryApplicationsAtomic2010a,boninElectricDipolePolarizabilitiesAtoms1997}, 
the intermediate result of Eq.~\ref{eq:E002int} is then 
\begin{align}
    \frac{\delta A_{g,E}}{A_g} 
    \approx - \left( \frac{E^2}{2 \hat{E}} \right) \left(  \frac{4 e^2 \Braket{ r_a^2 }}{3 E_a} + \alpha_a(0) \right) 
    \approx - \left( \frac{E^2}{2 \hat{E}} \right) \left(  \frac{4 e^2 \Braket{ r^2 }}{3 E_a} + \alpha_a(0) \right)\,. 
    \label{eq:dA_E} 
\end{align} 
The resulting hyperfine shift is usually summarized with an isotope-dependent Stark shift coefficient, 
$k = \delta A_{g,E} (I+\nicefrac{1}{2})/(h E^2)$, 
which is precisely known for many group-1 atoms due to its connection to BBR shifts in microwave atomic clocks \cite{mitroyTheoryApplicationsAtomic2010a}.

What remains is to choose a value for the effective energy $\hat{E}$. 
Ref.~\onlinecite{mcguyerHyperfinefrequencyShiftsAlkalimetal2013} chose $E_P$ for this value, which is the energy difference between the $g$S ground state and first excited P state (neglecting fine structure), giving 
\begin{align}
    k
    \approx 
    -\frac{(I+\nicefrac{1}{2})A_g}{2hE_P}\left(\frac{4e^2\Braket{r^2}}{3E_a}+\alpha_a(0)\right)\,.
    \label{eq:k} 
\end{align} 
The motivation for this choice came from using the approximation 
$E_\mu-E_0 \approx E_P$ 
and closure 
with the static polarizability $\alpha_a(0) = - (2/E^2) \delta E_{00,2}^{(U_E)} $, 
giving $\alpha_a(0) \approx 2 e^2 \Braket{r_a^2}/(3 E_P) \approx 2 e^2 \Braket{r^2}/(3 E_P)$, which is accurate to within 5\% for Na--Cs \cite{safronovaRelativisticManybodyCalculations1999}. 

Note that this shift coefficient is of the form of the intermediate result of Eq.~\ref{eq:E002int}. 
Alternatively, the shift coefficient can be put in the form of Eqs.~\ref{eq:HFS} and \ref{eq:HFS_Eab}. 
Using Eq.~\ref{eq:DefineEab}, the energy 
$E_{ab} \approx 2 e^2 \Braket{r_a^2}/[3 \alpha_a(0)] \approx E_P$, which then gives 
\begin{align}
    k\approx 
    -\frac{(I+\nicefrac{1}{2})A_g\alpha_a(0)}{2h}\left(\frac{2}{E_a}+\frac{1}{E_{P}}\right)\,.
    \label{eq:k_alt}
\end{align}
%
We estimate the coefficients $k$ using both Eqs.~\ref{eq:k} and \ref{eq:k_alt} in Section~\ref{ssec:r_stark} for a representative isotope of each group-1 atom.

\section{\label{sec:methods}Methods}

We computed the characteristic energies $E_a$ for each group-1 atom, and used these energies to estimate the resulting hyperfine shifts for the three applications described above. 
For convenience, Table~\ref{TableRef} in the Appendix provides many properties of the atoms used in these computations. 
As summarized below, the computation of $E_a$ and its uncertainty followed the numerical approach of Ref.~\onlinecite{mcguyerHyperfinefrequencyShiftsAlkalimetal2013} used previously for Na, K, Rb, and Cs. 
We revised this approach to provide updated values for those atoms, and extended it to provide values for the remaining group-1 atoms H, Li, and Fr. 
Additionally, to check these numerical results, we estimated $E_a$ using a scaling approach. 
For non-relativistic H with no QED corrections (or ``textbook'' H), this approach is exact and provides a check of the numerical result. 
For the other atoms, this approach is approximate and comes from adapting the calculation for H, providing a rough check of the similarity noticed between the numerical results and $2E_S$, where $E_S=E_{(g+1)}-E_g$ is the splitting between the ground and first excited $n$S states.

\subsection{\label{sec:m_numerics}Numerical Calculation} 

To numerically estimate the characteristic energy $E_a$, we calculated the final expression in Eq.~\ref{eq:Ea}. Our approach followed that outlined in Ref.~\onlinecite{mcguyerHyperfinefrequencyShiftsAlkalimetal2013}, which estimated $E_a$ for Na, K, Rb, and Cs, but not for H, Li, or Fr. 
This approach is semi-empirical in that it uses experimental values where available and extrapolates otherwise, as described below. 

For the $n$S-state energies $E_n$ and the ionization energies $E_\infty$, we used the values available in Ref.~\onlinecite{AtomicSpectraDatabase2009}. 
For the coupling coefficients $A_n$ of $^{1}$H, we used the experimental values for $A_1$ and $A_2$ from Refs.~\onlinecite{karshenboimHyperfineStructureHydrogen2002,bullisRamseySpectroscopy22023} and the theoretical values for $A_3$ to $A_8$ from Ref.~\onlinecite{jentschuraQuantumElectrodynamicCorrections2006}. 
For the coefficients $A_n$ of the remaining atoms, 
we used the experimental values for 
$^{7}$Li, $^{23}$Na, $^{39}$K, $^{87}$Rb, $^{133}$Cs, $^{210}$Fr, and $^{212}$Fr available in Ref.~\onlinecite{allegriniSurveyHyperfineStructure2022a}, which provides more values for Na, K, Rb, and Cs than were previously used by Ref.~\onlinecite{mcguyerHyperfinefrequencyShiftsAlkalimetal2013}. 
For Fr, we combined the values of two isotopes by using those for $^{212}$Fr directly and adding additional values from $^{210}$Fr after rescaling them using $A_g$. 
For K, we excluded $A_{11}$ and above because of their large uncertainties. 
For Rb and Cs, we excluded new values at very high $n$ where corresponding values of $E_n$ were unavailable. 
For Rb, we used the value of $A_6$ from Ref.~\onlinecite{ayachitulaPrecisionMeasurementHyperfine2024}. 

To extrapolate $E_n$ to higher $n$, we computed their corresponding effective quantum numbers 
\begin{align} 
n^*=\sqrt{R_\infty/(E_\infty-E_{n})}, 
\end{align} 
where $R_\infty$ is the Rydberg constant. 
These effective quantum numbers are expected to vary as 
\begin{align} \label{eq:nStarScaling} 
n^* \approx n - \delta_0\,. 
\end{align} 
This expression is exact for textbook H, with S-state quantum defect $\delta_0 = 0$ \cite{happerOpticallyPumpedAtoms2010}. 
Therefore, we used a linear fit of $n^*$ vs.\ $n$ to extrapolate both $n^*$ and $E_n$. 
The fitted values for $\delta_0$ closely matched the expected values of 
\begin{align} \label{eq:QuantumDefect} 
\delta_0 = g - \sqrt{R_\infty/E_\infty}. 
\end{align} 

Similarly, the coefficients $A_n$ are expected to vary as 
\begin{align} \label{eq:AnScaling} 
A_n \propto (n^*)^{-3}, 
\end{align} 
which is exact for textbook H \cite{arimondoExperimentalDeterminationsHyperfine1977a}. 
Therefore, we used a linear fit of ln$(A_n)$ vs.\ ln$(n^*)$ to extrapolate $A_n$. 
The fitted slope closely matched the expected value of $-3$ for all atoms except Li, for which $A_6$ and $A_7$ were outliers from this scaling. 
We adjusted the fit coefficients and uncertainties for Li to capture the influence of these outliers. 

To calculate the radial matrix elements, 
we used real-valued, reduced radial wave functions $P_{n0}(r)$, 
which relate to the spatial wave functions of the $n$S valence electron as 
\begin{align}
    \psi_{n\text{S}}(\mathbf r) 
        = \frac{P_{n0}(r)}{2 \sqrt{\pi} \, r}. 
\end{align}
Here, and subsequently, let $r$ be in units of $a_0$, such that we can set $a_0 \rightarrow 1$. 
Using these radial functions, the matrix elements are equivalent to the integrals 
\begin{align}
    \Braket{n\text{S}|r^2|g\text{S}} = \int_0^\infty P_{n0}(r) P_{g0}(r)\, r^2\,dr. 
    \label{eq:matrixelements}
\end{align}
These functions are zero at the origin, $P_{n0}(0) = 0$, and their slopes give the nuclear value as $\psi_{n\text{S}}(0) = (2 \sqrt{\pi})^{-1} dP_{n0}(0)/dr$. 
To match the sign convention of the square-root formula, these nuclear values must be positive, 
$\psi_{n\text{S}}(0) > 0$, 
which leads to negative values for all off-diagonal elements ($n \neq g$) calculated using Eq.~\ref{eq:matrixelements}. 

To compute these integrals for H, we used the exact, non-relativistic, wave functions \cite{griffithsIntroductionQuantumMechanics2018},
\begin{align}   \label{eq:Pn0H} 
P_{n0}(r) = \frac{2 r e^{-r/n}}{n^{5/2}} \,  L_{n-1}^1 \left( \frac{2r}{n }\right) , 
\end{align} 
where $L_n^k(x)$ is an associated Laguerre polynomial. 
For the ground state, 
\begin{align} \label{P10H}
P_{10}(r) = 2 r e^{-r/n} \, , 
\end{align} 
which gives $\Braket{r^2} = 3 a_0^2=3$ for H. 
For all other group-1 atoms, 
we used tabulated Roothaan-Hartree-Fock (RHF) wave functions and ground-state expectations $\Braket{r^2}$ (triple-zeta form if available) \cite{bungeRoothaanHartreeFockGroundStateAtomic1993, mcleanROOTHAANHARTREEFOCKATOMICWAVE}, 
and constructed Coulomb-approximation (CA) wave functions for the excited states \cite{oretoBuffergasinducedShiftBroadening2004}. 
In the CA, the alkali-metal radial wave functions are proportional to Whittaker $W$ functions \cite{nationalinstituteofstandardsandtechnologyNISTHandbookMathematical2010} up to normalization, 
$P_{n0}(r) \propto W_{n^*,1/2}(2 r / n^*)$, 
which are given by the asymptotic series 
\begin{align}
    P_{n0}(r)\approx\sum_{q=0}^p c_q e^{-r/n^*} r^{n^*-q}\,,
\end{align} 
where the coefficients $c_q$ obey the recurrence relation $c_q=c_{q-1}\, n^*(n^*-q)(n^*-q+1)/2q$ \cite{oretoBuffergasinducedShiftBroadening2004}. 
We chose the upper limit $p$ to give the best convergence of the series at $r=1a_0=1$ when $E_n$ was available, and otherwise extrapolated $p$ by noting that it was very nearly equal to $n$ plus a constant offset for large $n$. 
To normalize the CA functions, we used a lower bound of $0.1 a_0$, such that 
$\int_{0.1}^\infty [P_{n0}(r)]^2\,dr = 1$. 
To account for the nuclear sign, we used the property 
$\lim_{r \rightarrow 0} \text{sgn}[P_{n0}(r)] = (-1)^{n+g} \lim_{r \rightarrow \infty} \text{sgn}[P_{n0}(r)] $ of the CA functions. 

Using these radial functions, we calculated the off-diagonal matrix elements of Eq.~\ref{eq:matrixelements} up to $n = 35$. 
To extrapolate these elements 
to higher $n$, we used a linear fit to the asymptotic behavior of $\ln\left(-\braket{g\text{S}|r^2|n\text{S}}\right)$ vs.\ $\ln(n^*)$ within the region of $n \in [28,35]$. 
For more on the scaling of these elements, see the next section.  

Together, using these experimental, calculated, and extrapolated values, we computed the energy $E_a$ for each group-1 atom by summing terms up to $n = 500$. 
To calculate uncertainties for these values, we numerically estimated the contributions from all fitted extrapolations, from the radial integration bounds, and from the non-orthogonality of the CA and RHF functions. 
The uncertainties also include contributions from the square-root formula \cite{bouchiatMagneticDipoleElectric1988,dzubaOffdiagonalHyperfineInteraction2000}, the isotope dependence of $A_n$ (except for H) \cite{perssonTableHyperfineAnomaly2023}, and the experimental uncertainty of $A_n$ \cite{allegriniSurveyHyperfineStructure2022a}. 
However, these uncertainties do not account for approximations in the derivations of formulas with $E_a$ for the three perturbations. 

\subsection{\label{sec:m_scaling}Scaling Estimation}

We analytically calculated the characteristic energy $E_a$ by computing the dimensionless ratio $\mathcal R = E_a / 2 E_S$, which using Eq.~\ref{eq:Ea}, is given by the sum 
\begin{align} \label{eq:ScalingRatioR} 
\frac{1}{\mathcal{R}} 
    = \frac{2 E_S}{E_a} 
    \approx 2\sum_{n>g}^\infty \, \sqrt{\left|\frac{A_n}{A_g}\right|} 
    \left(\frac{E_S}{E_g-E_n} \right)
    \frac{\Braket{n\text{S}|r^2|g\text{S}} }{\Braket{g\text{S}|r^2|g\text{S}}}\,. 
\end{align}
Each term in the sum is the product of three dimensionless factors. 
The first two factors have expected scalings given previously, which are exact for textbook H. 
Using Eq.~\ref{eq:AnScaling}, 
the first factor is 
\begin{align}  \label{eq:AnTermScaling}
\sqrt{\left|\frac{A_n}{A_g}\right|} 
    \approx \left( \frac{g^*}{n^*}\right)^{3/2}, 
\end{align}
where the effective quantum number for the ground state is $g^* = \sqrt{R_\infty/E_\infty}$. 
An optional improvement that gives slightly better agreement for the heavy alkali metals is to keep the first term, which gives 
$\sqrt{|A_n/A_g|} \approx \sqrt{|A_{g+1}/A_g|}\,[(g+1)^*/n^*]^{3/2}$. 
Using Eqs.~\ref{eq:nStarScaling} and \ref{eq:QuantumDefect}, 
the second factor is 
\begin{align}   \label{eq:EnTermScaling} 
    \frac{E_S}{E_g-E_n}= -\frac{E_{g+1}-E_g}{E_n-E_g}\approx-\frac{1-[g^*/(g+1)^*]^2}{1-(g^*/n^*)^2}\,.
\end{align}

The remaining third term of radial matrix elements can be calculated exactly for textbook H. 
Using the radial functions of Eq.~\ref{eq:Pn0H}, the off-diagonal elements are 
\begin{align}
    \Braket{n\text{S}|r^2|g\text{S}} 
    &=\int_{0}^\infty P_{10}(r)\,P_{n0}(r)\,r^2\,dr
    =\frac{n^{5/2}}{8}\int_{0}^\infty z^4\,e^{-z(n+1)/2}\,L_{n-1}^1\left(z\right)\,dz\,, 
\end{align}
where $z=2r/n$. 
A more general form of this integral is given in Eq.~7.414(7) of Ref.~\onlinecite{gradstejnTableIntegralsSeries2009}: 
\begin{align}
    \int_0^\infty z^\beta\, e^{-s z}\, L_{n}^1(z)\, dz 
	&= \frac{ \Gamma(\beta + 1)\Gamma(n + 2)}{ n!\, s^{\beta + 1}} \,{}_2F_1(-n,\beta+1; 2; 1/s) \,,
\end{align}
if $\Re(\beta)>0$ and $\Re(s)>0$, 
where $_2F_1$ is the ordinary hypergeometric function. 
When $n$ is a positive integer, this gives 
\begin{align} 
\int_0^\infty z^4 e^{-z(n+1)/2} L_{n-1}^1(z) dz 
	&= - \frac{ 256 n^2 }{ (n^2-1)^3 } \left(\frac{n-1}{n+1} \right)^n \,. 
\end{align} 
Combining results, and noting that $\Braket{r^2} = 3a_0=3$, this gives the off-diagonal elements for H as 
\begin{align}   \label{eq:Hradialscaling} 
\frac{ \Braket{ n \text{S} | r^2 | g \text{S} } }{ \Braket{ g \text{S} | r^2 | g \text{S} } }
	&= - \frac{32 n^{9/2} }{ 3 (n^2-1)^3 }  \left(\frac{n-1}{n+1} \right)^n 
    \qquad (n \neq g)\,. 
\end{align} 

For the alkali metals, while the integrals of approximate radial functions can be evaluated analytically, the resulting expressions are unwieldy. 
Instead, we proceeded by adapting the result for H to use effective quantum numbers as follows. 
For the excited state, we directly substituted $n \rightarrow n^*$. 
For the ground state, however, a direct substitution is not straightforward, since $g = 1$ for H. 
To proceed, we appealed to how different CA/hydrogenic radial matrix elements are often connected by an energy difference factor (c.f.\ Eq.~9 of Ref.~\onlinecite{sanchezMatrixelementCalculationsHydrogenlike1992}) and tried substitutions of the form 
$E_n - E_g \propto (n)^{-2} - (g)^{-2} \rightarrow (n^*)^{-2} - (g^*)^{-2}$, or equivalently, 
$(n^{2} - 1)/n^2 \rightarrow [(n^*)^2 - (g^*)^2]/(n^* g^*)^2$. 
We found that applying this substitution to the factor $(n^2-1)^3$ in the denominator of Eq.~\ref{eq:Hradialscaling} best adapted the relative variation of the elements vs.\ $n^*$ to the alkali metals. 
Next, we noticed that the numerical results for 
${ \Braket{ (g+1) \text{S} | r^2 | g \text{S} } }/{ \Braket{ g \text{S} | r^2 | g \text{S} } } \approx - 0.7957 \pm 1.5$\%, with the exception of Fr, which seems to be an outlier with a value closer to $-0.90$. 
Therefore, we normalized that first ratio to the average value of about $-0.7957$, which gave  
\begin{align}  \label{eq:AlkaliRadialScaling} 
\frac{ \Braket{ n\text{S} | r^2 | g\text{S} } }{ \Braket{ g\text{S} | r^2 | g\text{S} } } 
	&\approx - 0.7957 \left(\frac{n^*}{g^* + 1} \right)^{9/2} \left( \frac{ 2 g^* + 1 }{ (n^*)^{2}-(g^*)^{2} } \right)^3  \left(\frac{n^*-1}{n^*+1} \right)^{n^*}  \left(\frac{g^* + 2}{g^*} \right)^{g^*+1} 
    \qquad (n \neq g)\,. 
\end{align} 

Finally, using Eqs.~\ref{eq:AnTermScaling}, \ref{eq:EnTermScaling}, and either \ref{eq:Hradialscaling} or \ref{eq:AlkaliRadialScaling} with Eq.~\ref{eq:ScalingRatioR} gives expressions for the ratio $\mathcal{R}$ that are solely functions of $g^*$. 
For textbook H, the exact ratio simplifies to 
\begin{align} \label{eq:ScalingH}
    \frac{1}{\mathcal{R}}  
	&\approx  
	16 \sum_{n = 2}^\infty  \frac{n^5 (n-1)^{n-4}}{(n+1)^{n+4}}  
    \qquad (g^* = 1)\,, 
\end{align}
and for the alkali metals, the approximate ratio simplifies to 
\begin{align} \label{eq:ScalingAlkali}
\frac{1}{\mathcal{R}} 
	&\approx  
	\left[1.5914 \, 
	\frac{(2 g^* + 1)^4}{(g^*+1)^2}
	\left(\frac{g^* + 2}{g^*} \right)^{g^*+1} 
	\frac{(g^*)^{3/2}}{(g^*+1)^{9/2}} \right] 
	\sum_{k = 1}^\infty 
	\frac{ (g^* + k)^5 }{ k^4 (2 g^* + k)^4 } 
	\left(\frac{g^*+k-1}{g^*+k+1} \right)^{g^* + k} 
    \qquad (g^* \neq 1)\,, 
\end{align}
where the summation index $k = n^* - g^* = n - g \geq 1$.

\section{\label{sec:results}Results}

\subsection{\label{ssec:Ea}Characteristic Energies}

Table~\ref{TableEa} summarizes our results for the characteristic energies $E_a$. 
Our main results are the numerically calculated values and uncertainties for each group-1 atom. 
The fractional uncertainty is lowest for Na, around 3\%, and highest for Fr, around 25\%.
In general, the heavier atom uncertainties are dominated by RHF-CA non-orthogonality, which increases with $g$.
This trend captures the estimated sensitivity to radial wave function details, from the greater number of core electrons. 
For Li, the uncertainty is dominated by an irregular trend in $A_n$ values, as discussed above, and for H, the dominant contributions are the square-root formula and sensitivity to numerical details.

Table~\ref{TableEa} also provides values of the dimensionless ratio $\mathcal R=E_a/2E_S$. 
The ratios using numerically calculated values for $E_a$ agree well with values computed using the exact scaling of Eq.~\ref{eq:ScalingH} for H and approximate scaling of Eq.~\ref{eq:ScalingAlkali} for the alkalis.
The scaling result for Li agrees worst with the numerical result, which is expected, given its aforementioned irregular trend of $A_n$ vs.\ $n^*$.
Plots of various quantities comparing numerical and scaling results generated in our code confirm this general pattern of agreement. 
These plots are included in an online repository, along with our code \cite{Note1}.
A heuristic argument for this agreement with the alkali metals is that in the sum of $1/\mathcal R$, the first term contributes about 80\% of the total value. 
In that first term, the energy denominator is $E_S$, $\sqrt{|A_n/A_g|}\approx 0.5$ for Li--Fr, and the radial ratio is about $0.80$. 
Together, completing the tail sum removes that 0.80 term, leaving $2E_S$ in the denominator. 
Interestingly, numerically calculating $E_a$ using only CA wave functions (not shown) caused the ratio $\mathcal R$ to follow an increasing trend up to $\mathcal R\approx 1.25$ for Fr, highlighting its sensitivity to radial wave function details.

For Na--Cs, the updated physical parameters, extrapolations, and minor numerical improvements did not significantly change the results from those previously reported in Ref.~\onlinecite{mcguyerHyperfinefrequencyShiftsAlkalimetal2013} (and reproduced in the table). 
Our updated calculation for Na has smaller uncertainty than the previous result due to improved extrapolation in $p$ for the CA wave functions.
For the other alkali metals, we revised our estimate for the uncertainty due to CA--RHF non-orthogonality, to better capture sensitivity to  radial function details, which slightly increased the overall uncertainties. 
In all cases, the numerical results disagree with a previous estimates for $E_a\approx \overline E_{e}$ given in Ref.~\onlinecite{vanierQuantumPhysicsAtomic1989} (and reproduced in the table) by  up to a factor of roughly 2.
They also differ from the previous estimate $E_a\approx E_\infty$ of Ref.~\onlinecite{hermanFrequencyShiftsHyperfine1961} (reproduced in the table), especially for lighter atoms.

\begin{table}[htb] 
\caption{\label{TableEa} 
Characteristic energies $E_a$ (eV) of Eqs.~\ref{eq:HFS}, \ref{eq:Ea2ndOrderResult}, \& \ref{eq:HFS_Eab} estimated using Eq.~\ref{eq:Ea}. 
The resulting numerically calculated ratios $E_a/2E_S\approx 1$, which we confirmed using scaling argument as shown. 
We include previously suggested values for $E_a$ for comparison. 
The values in parentheses are uncertainties in the last digits. 
}
\begin{ruledtabular}
\begin{tabular}{l | c c c c c c c }
Group-1 atom 	
	& H 			& Li 			& Na 		& K 			& Rb 		& Cs 		& Fr \\
\hline
$E_a$ (this work)  \\ 
\qquad Numerical calculation 
	& 22.2(8) 		& 6.53(33) 	& 6.48(20) 	& 5.29(34) 	& 5.06(46) 	& 4.58(60) 	& 4.8(1.1) \\ 
\qquad Ratio $\mathcal R = E_a/(2 E_S)$ \\ 
\qquad\qquad Numerical calculation 
	& 1.09(4) 	& 0.97(5) 	& 1.02(3) 	& 1.01(7) 	& 1.01(9) 	& 1.00(13) 		& 1.0(2) \\ 
\qquad\qquad Scaling (exact, Eq.~\ref{eq:ScalingH}) 
	& 1.087 \\ 
\qquad\qquad Scaling (approximate, Eq.~\ref{eq:ScalingAlkali}) 
	&  			&  1.038	& 1.026 	& 0.986 	& 0.977 	& 0.962 	& 0.972 \\ 
$E_a$ (previous work, \cite{mcguyerHyperfinefrequencyShiftsAlkalimetal2013}) \\ 
\qquad Numerical calculation 
	& 			& 			& 6.55(33)		& 5.31(28)		& 5.05(37)		& 4.59(48) \\		
Previously suggested values for $E_a$ \\ 
\qquad $\overline{E}_e$ of Ref.~\onlinecite{vanierQuantumPhysicsAtomic1989}  
	& 11.90 		& 3.62 		& 3.62		& 2.98		& 2.88 		& 2.66 \\ 
\qquad $E_\infty$~\cite{AtomicSpectraDatabase2009} following 
Ref.~\onlinecite{hermanFrequencyShiftsHyperfine1961}  
	& 13.598 		& 5.392		& 5.139		& 4.341		& 4.177 		& 3.894  		& 4.073 
\end{tabular}	
\end{ruledtabular}
\end{table}

\subsection{\label{ssec:surface}Wall Shifts Near Surfaces}

Table~\ref{TableSurface} summarizes calculations of the hyperfine shifts for group-1 atoms due to interaction with a conducting surface at a chosen distance of $R=0.1~\mu$m, computed using Eq.~\ref{eq:HFS_wall}.
The uncertainty is dominated by the uncertainty in $E_a$, which generally increases with mass.
The predicted shifts of a representative clock frequency, $\delta f_\text{clock}=(I+\nicefrac{1}{2})\delta A_g/h$, fall in the Hz to tens of Hz range for this distance. 
Such shifts are resolvable with current techniques, and would be even more so at closer distances $R$, since they scale with $R^{-3}$. 
Note that $^{212}$Fr has no ``0--0'' clock transition; the listed value is representative of the scale of ground-state hyperfine shifts.

\begin{table}[ht] 
\caption{\label{TableSurface}
Hyperfine shifts of group-1 atoms near conductive surfaces. 
Values were calculated using Eq.~\ref{eq:HFS_wall} 
with group-1 atom-surface $C_3$ coefficients from Ref.~\onlinecite{dereviankoElectricDipolePolarizabilities2010a}
and with a distance of $R = 0.1~\mu$m from the surface. 
Representative clock frequency shifts are $\delta f_\text{clock}=(I+\nicefrac{1}{2})\delta A_g/h$. 
}
\begin{ruledtabular}
\setlength{\tabcolsep}{2.5pt}
\begin{tabular}{l | c c c c c c c}
Atom & $\text{H}$ & $\text{Li}$ & $\text{Na}$ & $\text{K}$ & $\text{Rb}$ & $\text{Cs}$ & $\text{Fr}$ \\
\hline
\rule[0pt]{0pt}{2.5ex}$(\delta A_g/A_g)\times10^{9}$ & $-0.091(4)$ & $-1.87(10)$ & $-2.33(8)$ & $-4.4(3)$ & $-5.5(5)$ & $-7.5(1.0)$ &  $-7(2)$ \\ 
\hline 
\hline
\rule[0pt]{0pt}{2.5ex}Isotope 	
	& $^1$H 		& $^{7}$Li 	& $^{23}$Na 	& $^{39}$K 	& $^{87}$Rb 	& $^{133}$Cs 	& $^{212}$Fr \\
\hline 
\rule[0pt]{0pt}{2.5ex}$\delta f_\text{clock}$~(Hz)   & $-0.129(5)$ & $-1.50(8)$ & $-4.13(13)$ & $-2.04(13)$ & $-37(4)$ & $-52(7)$ & $-372(90)$ \\ 
\end{tabular}	
\end{ruledtabular}
\end{table}

\subsection{\label{ssec:Eab}Pressure Shifts During Collisions}

Table~\ref{TableEab} summarizes values of the correction energies $E_{ab}$ for collisions between group-1 atoms and group-18 (noble gas) atoms. 
These computed values all satisfy $E_{ab} > E_a/2$, and 
the $E_a$ term usually dominates over the $E_{ab}$ term in Eq.~\ref{eq:HFS}, excluding H. 
The listed uncertainties are due to the uncertainties of $C_3$ and $C_6$ from Ref.~\onlinecite{dereviankoElectricDipolePolarizabilities2010a}.

\begin{table}[ht] 
\caption{\label{TableEab} 
Energies $E_{ab}$ of expressions \ref{eq:HFS} \& \ref{eq:HFS_Eab} for group-1--group-18 (noble gas) collision pairs. 
Values were computed using Eq.~\ref{eq:EabCollision} with values of $C_6$ from Table G of Ref.~\onlinecite{dereviankoElectricDipolePolarizabilities2010a}, of $C_3$ from Table H of Ref.~\onlinecite{dereviankoElectricDipolePolarizabilities2010a}, 
and of RHF matrix elements $\braket{r^2}$ from Refs.~\onlinecite{bungeRoothaanHartreeFockGroundStateAtomic1993,mcleanROOTHAANHARTREEFOCKATOMICWAVE}.
Uncertainties are solely due to $C_3$ and $C_6$, where available. 
}
\begin{ruledtabular}
\begin{tabular}{l | c c c c c c c c }
	
$E_{ab}$ (eV) & He 	& Ne 	& Ar 	 	& Kr 		& Xe \\ 
\hline  
\rule[0pt]{0pt}{2.5ex}\qquad H   & $43.546$ & $55.0(3)$  & $36.15(19)$  & $35.3(3)$ & $33.02(16)$   \\ 
\qquad Li  & $32.37(3)$     & $42.1(2)$ & $24.46(15)$ & $22.99(9)$  & $20.43(10)$  \\ 
\qquad Na & $33.23(4)$     & $43.0(3)$   & $25.33(13)$ & $23.88(17)$ & $21.30(10)$  \\ 
\qquad K  & $33.24(7)$     & $42.9(3)$   & $25.42(18)$ & $24.06(17)$ & $21.47(13)$  \\ 
\qquad Rb  & $33.61(9)$     & $43.2(3)$   & $25.84(16)$ & $24.58(15)$ & $21.96(12)$  \\ 
\qquad Cs & $34.2(2)$    & $43.9(4)$   & $26.4(2)$ & $25.14(17)$ & $22.54(17)$  \\ 
\qquad Fr & $38.4(3)$      & $49.3(5)$   & $30.0(4)$   & $28.7(3)$   & $25.8(3)$    
\end{tabular}	
\end{ruledtabular}
\end{table}

\subsection{\label{ssec:r_stark}Stark Shifts in Static Electric Fields}

Table~\ref{TableK} summarizes calculations of the Stark-shift coefficients $k$ computed using Eqs.~\ref{eq:k} and \ref{eq:k_alt}. 
The listed uncertainties are due to $E_a$ alone. 
The values computed using Eq.~\ref{eq:k} all agree, as expected, with previous calculations using the same expression in Ref.~\onlinecite{mcguyerHyperfinefrequencyShiftsAlkalimetal2013} and, with the exception of H, with experimental results from Refs.~\onlinecite{mitroyTheoryApplicationsAtomic2010a,stuartDifferentialStarkShifts1980a}. 
The values computed using Eq.~\ref{eq:k_alt} are comparable, and more importantly, also agree with experimental results, up to the same exception for H. 
Therefore, both forms for $k$ are equivalent with respect to the available experimental values and their uncertainties. 
For the case of H, the mutual disagreement likely stems from neglecting a contribution from continuum states. 
For H, the discrepancy between our result for $k$ and the measured value matches the roughly $20$\%  contribution to $\alpha_a(0)$ from the continuum that we neglect, mentioned by Ref.~\onlinecite{mitroyTheoryApplicationsAtomic2010a}, compared to approximately 10\% for Cs, mentioned by Ref.~\onlinecite{angstmannFrequencyShiftCesium2006}. 
For each computed value of $k$, the fractional contribution of the $E_a$ term in Eqs.~\ref{eq:k} or \ref{eq:k_alt} was roughly 40\%. 
For Fr, there is no experimental value available, so for comparison, we scaled a semi-empirical result computed for $^{210}$Fr from Ref.~\onlinecite{aokiQuantumSensingElectron2021} by $(I+\nicefrac{1}{2})A_g$, and the resulting value is consistent with our prediction. 

\begin{table}[htb] 
\caption{\label{TableK}
Stark-shift coefficients $k$ ($10^{-10}$ Hz/(V/m)$^{2}$) estimated using Eqs.~\ref{eq:k} and \ref{eq:k_alt}. 
The values in parentheses are uncertainties in the last digits, which for this work only include the contributions due to $E_a$. 
Experimental values from Refs.~\onlinecite{mitroyTheoryApplicationsAtomic2010a,stuartDifferentialStarkShifts1980a} are included for comparison, as well as a semi-empirical value for $^{210}$Fr scaled by $A_g(I+\nicefrac{1}{2})$ from Ref.~\onlinecite{aokiQuantumSensingElectron2021}. 
}
\begin{ruledtabular}
\begin{tabular}{l | c c c c c c c }
Atom 	& $^{1}$H & 	$^{7}$Li 			& $^{23}$Na		& $^{39}$K 		& $^{87}$Rb 		& $^{133}$Cs 	& $^{212}$Fr \\
\hline
\rule[0pt]{0pt}{2.5ex}$k$ (this work)  \\ 
\qquad Value (using Eq.~\ref{eq:k}): 
		& $-$0.000675(12) & $-$0.0588(11)	& $-$0.121(2)		& $-$0.075(2)		& $-$1.29(5)		& $-$2.50(15) & $-$10.7(1.3) \\ 
\qquad Value (using Eq.~\ref{eq:k_alt}): 
		& $-$0.000620(10) & $-$0.0575(11)	& $-$0.1163(14)	& $-$0.069(2)	& $-$1.15(4)		& $-$2.15(11) & $-$8.3(8)\\ 
$k$ (previous work, \cite{mcguyerHyperfinefrequencyShiftsAlkalimetal2013}) 
		& 		& 					& $-$0.120(3)		& $-$0.074(2)		& $-$1.29(4) 		& $-$2.50(13) \\
$k$ (experimental, \cite{mitroyTheoryApplicationsAtomic2010a,stuartDifferentialStarkShifts1980a}) 
		& $-$0.00084(2) 
					& $-$0.061(2) 		& $-$0.124(3)		& $-$0.071(2)		& $-$1.23(3) 		& $-$2.271(4) 
                    &  \\
$k$ (semi-empirical, \cite{aokiQuantumSensingElectron2021}) & & & & & & & $-8.47 $
                    
\end{tabular}	
\end{ruledtabular}
\end{table}

\section{\label{sec:discussion}Discussion}

In the previous section, we reported calculations of several experimentally-relevant parameters for group-1 atoms perturbed by collisions with group-18 atoms, by static electric fields, or by nearby conducting surfaces.
We calculated updated or extended values for the characteristic energy $E_a$, which appears in ground-state hyperfine shifts in all three of the applications derived above, and likely  others.
These updated results for $E_a$ are based on numerical sums using experimental measurements where available. 
Using an exact scaling calculation in the case of textbook H, we validated our numerical approach using the principal-quantum-number dependence of some atomic properties.
We adapted this scaling method, using some approximations,  to validate the numerical result for the alkali metals, and found good agreement between the two approaches.

Compared to older estimates for the characteristic energy $E_a$, our results are larger by up to a factor of about 2.
For Na--Cs, our results are very close to our prior calculations using the same method \cite{mcguyerHyperfinefrequencyShiftsAlkalimetal2013}.
Updated inputs and slight numerical improvements only led to small changes in their values;
at the level of estimated uncertainty, there were no significant changes in $E_a$ for Na--Cs. 

One surprising result from both our numerical calculations and scaling arguments is that the characteristic energy $E_a$ is close to  $2E_S$.
In particular, the ratio $E_a/2E_S$ was very nearly unity for all of the alkali metals. 
This coincidence merits further investigation. 
From exploring how the ratio changed when using only CA functions instead of CA with RHF functions, we speculate that the fine details of the radial wave functions play an important role in this coincidence.

For long-range collision shifts, our updated and expanded results for $E_{a}$ and $E_{ab}$ should prove useful to other work investigating pressure shifts and interaction potentials between group-1 and group-18 atoms, such as  Refs.~\onlinecite{ishikawaCollisionalShiftsHyperfine2024,ishikawaFlyingCharacterizationColliding2023,ishikawaNoblegasAtomsCharacterized2022,ishikawaPseudopotentialAnalysisHyperfine2023}. 
They could also potentially have implications for experiments with cold atoms, since the long-range van der Waals (vdW) potential influences scattering in the vicinity of Feshbach Resonances \cite{Chin2010}, as well as in magnetoassociation \cite{jones_ultracold_2006,zuchowskiUltracoldRbSrMolecules2010, brueMagneticallyTunableFeshbach2012}.
Note that to calculate $E_{ab}$ for this application, we had to choose particular values of $C_3$ and $C_6$, which influences the values and uncertainty of $E_{ab}$. 
For example, we used the values of $C_6$ reported in Ref.~\onlinecite{dereviankoElectricDipolePolarizabilities2010a}, which have $\lesssim 1\%$ fractional uncertainty. 
Some other methods for evaluating $C_6$ agree with Ref.~\onlinecite{dereviankoElectricDipolePolarizabilities2010a} within their estimated uncertainties \cite{zhangLongrangeDispersionInteractions2017a}, but others differ by up to 5\% (with no estimated uncertainties) \cite{eshelRoleAdiabaticityControlling2017}. 

For the Stark shift coefficients $k$, we found good agreement with experimental results for the alkali metals using both the intermediate form of Eq.~\ref{eq:E002int}, which follows our previous work \cite{mcguyerHyperfinefrequencyShiftsAlkalimetal2013}, and the final form of Eq.~\ref{eq:k_alt} that we derived to better match Eq.~\ref{eq:HFS}, which used $E_{ab} \approx E_P$. 
At the level of available uncertainty, these two forms are indistinguishable compared to experiment. 
The agreement using Eq.~\ref{eq:k_alt} is sensitive to the exact choice of $E_{ab}$: 
The value of $E_P$ came from using $E_{ab} \approx 2 e^2 \Braket{r_a^2}/[3 \alpha_a(0)]$ and then $\Braket{r_a^2} \approx \Braket{r^2}$. 
Interestingly, for all atoms, the agreement slightly worsened using $E_{ab} \approx 2 e^2 \Braket{r_a^2}/[3 \alpha_a(0)]$ directly, and even more so using Eq.~\ref{eq:C3Trick} to replace $\Braket{r_a^2}$ with an atom-surface $C_3$ for the group-1 atom. 

A testable prediction of our calculations is the wall shift for group-1 atoms near conducting surfaces. 
While our derivation assumed a perfect conductor, similar atom-surface interactions near arbitrary surfaces can be modeled using an effective dielectric constant \cite{wylieQuantumElectrodynamicsInterface1984,fichetVanWaalsInteractions1995}.
Using clock transitions in alkali atoms, many of these shifts should be measurable using current techniques for achievable wall distances of $R\approx 0.1~\mu$m \cite{ballandQuectonewtonLocalForce2024}.
The typical atom--surface distance in chip traps is much larger, $R\approx 10$--$100~\mu$m \cite{fortaghMagneticMicrotrapsUltracold2007}, and some optical lattice setups trap atoms as close as $R\approx 1~\mu$m from a window \cite{grossQuantumGasMicroscopy2021}, so these shifts should be negligible for most  cold-atom experiments.
In hollow-core fiber applications, alkali atoms are closer to the fiber surfaces, typically $R\approx 0.1$--$10~\mu$m, though the wall shift predicted here would be dominated by other effects \cite{okabaLambDickeSpectroscopyAtoms2014}.
One emerging field where atoms are held much closer to surfaces is in nanostructured slot waveguides \cite{ritterCouplingThermalAtomic2018} or nano-grating experiments \cite{lecoffreMeasurementCasimirPolderInteraction2025}, where the atoms can be $R\approx 50$~nm from the surface.
Some research on alkali metals in nanocells probes wall interactions for $R\approx 50$--$250$~nm \cite{whittakerSpectroscopicDetectionAtomsurface2015,sargsyanFeaturesVanWaals2026,sargsyanStudyInteractionRubidium2023,sargsyanCompetingVanWaals2023,sargsyanDopplerfreeSpectroscopyCs,sargsyanStudyVanWaals2025}.
The response of an atom to a nearby surface can also be altered using surface plasmon resonances engineered to match the atom's transition frequency \cite{chanTailoringOpticalMetamaterials2018}. 
As suggested by Ref.~\onlinecite{araujoCooperativeAtomicEmission2024} for optical transitions, such a system could be used to control cooperative emission for quantum information or metrology.
A similar concept with hyperfine sublevels in the alkali atoms could employ spoof surface plasmons \cite{garcia-vidalSpoofSurfacePlasmon2022} to observe super- or sub-radiant microwave emission, and potentially improve the performance of microwave clocks, similar to \cite{norciaFrequencyMeasurementsSuperradiance2018}.

Our derivations here require only a modest set of conditions on the form of the interactions, so are fairly general.
A similar approach could be useful for extending this work to estimate shifts due to other interactions.
The most straightforward are other interactions where $U\propto \braket{r^2}$, but a similar approach could be used with other radial matrix elements, such as different radial powers and/or radial derivatives, likely leading to different versions of corresponding characteristic energies. 
Possible ways to explore this adaptation would be to modify our numerical code to calculate different elements, or to adapt the scaling relations by using commutator relations to convert them to model different radial elements using the coulomb approximation (c.f.\ Eq.~9 of Ref.~\onlinecite{sanchezMatrixelementCalculationsHydrogenlike1992}).

\section{\label{sec:conclusion}Conclusion}

In this article, we treated three scenarios that lead to shifts of the ground-state hyperfine coupling in group-1 atoms with the particular form of Eq.~\ref{eq:HFS}.
We showed how such shifts can arise as either a second-order effect (i.e., surface interactions) or as a third-order effect (i.e., collisions and external electric fields). 
In all cases, a characteristic energy $E_a$ appears that depends only on the properties of the group-1 atom. 
For third-order shifts, another energy $E_{ab}$ appears that corrects for additional details of the interaction (e.g., choice of collision partner).  
Using updated experimental values and improved numerical techniques, we calculated $E_a$, $E_{ab}$ for collisions with group-18 (noble gas) atoms, van der Waals shifts near surfaces, and Stark/BBR shift coefficients $k$ for static electric fields. 
We found good agreement with experimental results for $k$, where available, and provided a testable prediction of hyperfine shifts near surfaces. 

Our approach provides a straightforward way to estimate various experimentally relevant parameters using a small number of spectroscopic and calculated parameters.
This approach should extend to estimate hyperfine shifts for other interactions---our derivations of the shifts made fairly modest assumptions.
We expect future work will extend our approach to other interactions and applications.

\begin{acknowledgments}
B.\,A.\,O. acknowledges support from the M.\,J.\,Murdock Charitable Trust, and the National Science Foundation through Grant No.\,PHY-2418777.
\end{acknowledgments}

\section*{Data availability}
The data that support the findings of this article are openly available in the Zenodo repository at \href{https://doi.org/10.5281/zenodo.22149598}{10.5281/zenodo.22149598}, and an updated version is available in a  \href{https://github.com/olsenlab-science/group1-HFS}{github repository}.
The data include {\scshape matlab} code that computed the numerical results, example output (text and plots) produced by the code, and a {\scshape mathematica} notebook that computed scaling results. 

\appendix

\section{Reference values}

Table~\ref{TableRef} provides some reference values used in our calculations for the group-1 atoms.


\begin{table}[ht] 
\caption{\label{TableRef}
Reference values for group-1 atoms. 
Some uncertainties and full-precision values 
are suppressed for clarity, though calculations used additional precision where applicable. 
The ground-state principal quantum number $g$ and effective quantum number $g^*$ appear in expressions in Sec.~\ref{sec:m_scaling}.  
$E_D$ and $E_S$, and $E_\infty$ are the first excited P- and S-state term energies and the ionization energy, respectively. 
The value of $\langle g$S$ | r^2 | g$S$ \rangle$ for textbook H is exact, while the others are calculated using RHF approximate wave functions. 
a.u.~denotes atomic units. 
}
\begin{ruledtabular}
\begin{tabular}{l | c c c c c c c }
Atom 	
	& H 			& Li 			& Na 		& K 			& Rb 		& Cs 		& Fr \\
\hline
\rule[0pt]{0pt}{2.5ex}$g$ 	& 1			& 2			& 3			& 4			& 5			& 6 			& 7 \\ 
$g^* = \sqrt{R_\infty/E_\infty}$ 
	& 1 			& 1.5885		& 1.6271 		& 1.7704 		& 1.8048 		& 1.8693 		& 1.8278	 \\ 
$E_P$ (eV) \cite{AtomicSpectraDatabase2009} 
	& 10.1988		& 1.8478		 & 2.1037		&  1.6147 		& 1.5792 		& 1.4317 	 	& 1.6567 \\ 
$E_S$ (eV) \cite{AtomicSpectraDatabase2009}  
	& 10.1988 	& 3.3731 		& 3.1914 		& 2.6070 		& 2.4961		& 2.2981 	 	& 2.4474  \\ 
$E_\infty$ (eV) \cite{AtomicSpectraDatabase2009} 
	& 13.5984 	& 5.3917		& 5.1391		& 4.3407		& 4.1771 		& 3.8939  		& 4.0727 \\ 
$\langle g$S$ | (r/a_0)^2 | g$S$ \rangle$ \cite{bungeRoothaanHartreeFockGroundStateAtomic1993,mcleanROOTHAANHARTREEFOCKATOMICWAVE} 
	& 3		    & 17.7384		& 20.7048	 	& 31.5428		& 36.1679	 	& 44.9530	  	& 49.5441 \\ 
$\alpha_{a}(0)$ (a.u.) \cite{bayeExactNonrelativisticPolarizabilities2012,mitroyTheoryApplicationsAtomic2010a} 
	& 4.5 		& 164.2 		& 162.7 		& 290.8 		& 318.8  		& 401.0 		& 317.8 \\ 
$C_3$ atom-surface (a.u.) \cite{dereviankoElectricDipolePolarizabilities2010a} 
	& 0.25 		& 1.512 		& 1.871 		& 2.896 		& 3.426 		& 4.268 		& 4.437 \\ 
\hline 
\hline 
\rule[0pt]{0pt}{2.5ex}Isotope	
	& $^1$H 		& $^{7}$Li 	& $^{23}$Na 	& $^{39}$K 	& $^{87}$Rb 	& $^{133}$Cs 	& $^{212}$Fr \\
\hline 
\rule[0pt]{0pt}{2.5ex}$I$ 
	& $\nicefrac{1}{2}$			& $\nicefrac{3}{2}$			& $\nicefrac{3}{2}$				& $\nicefrac{3}{2}$				& $\nicefrac{3}{2}$				& $\nicefrac{7}{2}$	 		& 5 \\ 
$A_g/h$ (MHz) \cite{allegriniSurveyHyperfineStructure2022a}  
	& 1420.4058 	& 401.7520 	& 885.8131 	& 230.8599 	& 3417.3413 	& 2298.1579 	& 9064.2 \\ 
\end{tabular}	
\end{ruledtabular}
\end{table}

\bibliography{2026_HFS}

@article{adrianMatrixEffectsElectron1960,
  title = {Matrix {{Effects}} on the {{Electron Spin Resonance Spectra}} of {{Trapped Hydrogen Atoms}}},
  author = {Adrian, F. J.},
  year = 1960,
  month = apr,
  journal = {The Journal of Chemical Physics},
  volume = {32},
  number = {4},
  pages = {972--981},
  issn = {0021-9606, 1089-7690},
  doi = {10.1063/1.1730906},
  urldate = {2026-07-17},
  langid = {english}
}

@article{allegriniSurveyHyperfineStructure2022a,
  title = {Survey of {{Hyperfine Structure Measurements}} in {{Alkali Atoms}}},
  author = {Allegrini, Maria and Arimondo, Ennio and Orozco, Luis A.},
  year = 2022,
  month = dec,
  journal = {Journal of Physical and Chemical Reference Data},
  volume = {51},
  number = {4},
  pages = {043102},
  issn = {0047-2689, 1529-7845},
  doi = {10.1063/5.0098061},
  urldate = {2026-07-17},
  langid = {english}
}

@article{angstmannFrequencyShiftCesium2006,
  title = {Frequency {{Shift}} of the {{Cesium Clock Transition}} Due to {{Blackbody Radiation}}},
  author = {Angstmann, E. J. and Dzuba, V. A. and Flambaum, V. V.},
  year = 2006,
  month = jul,
  journal = {Physical Review Letters},
  volume = {97},
  number = {4},
  pages = {040802},
  issn = {0031-9007, 1079-7114},
  doi = {10.1103/PhysRevLett.97.040802},
  urldate = {2026-08-12},
  copyright = {http://link.aps.org/licenses/aps-default-license},
  langid = {english}
}

@article{aokiQuantumSensingElectron2021,
  title = {Quantum Sensing of the Electron Electric Dipole Moment Using Ultracold Entangled {{Fr}} Atoms},
  author = {Aoki, T and Sreekantham, R and Sahoo, B K and Arora, Bindiya and Kastberg, A and Sato, T and Ikeda, H and Okamoto, N and Torii, Y and Hayamizu, T and Nakamura, K and Nagase, S and Ohtsuka, M and Nagahama, H and Ozawa, N and Sato, M and Nakashita, T and Yamane, K and Tanaka, K S and Harada, K and Kawamura, H and Inoue, T and Uchiyama, A and Hatakeyama, A and Takamine, A and Ueno, H and Ichikawa, Y and Matsuda, Y and Haba, H and Sakemi, Y},
  year = 2021,
  month = oct,
  journal = {Quantum Science and Technology},
  volume = {6},
  number = {4},
  pages = {044008},
  issn = {2058-9565},
  doi = {10.1088/2058-9565/ac1b6a},
  urldate = {2026-08-07},
  langid = {english}
}

@article{araujoCooperativeAtomicEmission2024,
  title = {Cooperative Atomic Emission from a Line of Atoms Interacting with a Resonant Plane Surface},
  author = {Ara{\'u}jo, M. O. and Carvalho, J. C. De Aquino and Courteille, {\relax Ph}. W. and Laliotis, A.},
  year = 2024,
  month = sep,
  journal = {Physical Review A},
  volume = {110},
  number = {3},
  pages = {032813},
  issn = {2469-9926, 2469-9934},
  doi = {10.1103/PhysRevA.110.032813},
  urldate = {2026-06-29},
  langid = {english}
}

@article{arimondoExperimentalDeterminationsHyperfine1977a,
  title = {Experimental Determinations of the Hyperfine Structure in the Alkali Atoms},
  author = {Arimondo, E. and Inguscio, M. and Violino, P.},
  year = 1977,
  month = jan,
  journal = {Reviews of Modern Physics},
  volume = {49},
  number = {1},
  pages = {31--75},
  issn = {0034-6861},
  doi = {10.1103/RevModPhys.49.31},
  urldate = {2026-06-14},
  copyright = {http://link.aps.org/licenses/aps-default-license},
  langid = {english}
}

@misc{AtomicSpectraDatabase2009,
  title = {{{NIST Atomic Spectra Database}} ({{Version}} 5.12)},
  author = {Kramida, Alexander and Ralchenko, Yuri and Reader, Joseph and {NIST ASD Team}},
  year = 2009,
  month = jul,
  journal = {NIST Atomic Spectra Database},
  doi = {10.18434/T4W30F},
  urldate = {2026-07-14},
  langid = {english}
}

@article{ayachitulaPrecisionMeasurementHyperfine2024,
  title = {Precision Measurement of Hyperfine Constants and Isotope Shift of the {{Rb}} 6 {{S}} 1 / 2 State via a Two-Photon Transition},
  author = {Ayachitula, R. and Anderson, M. D. and McLaughlin, C. D. and Knize, R. J. and Mungan, C. E. and Lindsay, M. D.},
  year = 2024,
  month = aug,
  journal = {Physical Review A},
  volume = {110},
  number = {2},
  pages = {022803},
  issn = {2469-9926, 2469-9934},
  doi = {10.1103/PhysRevA.110.022803},
  urldate = {2026-07-28},
  langid = {english}
}

@article{ballandQuectonewtonLocalForce2024,
  title = {Quectonewton {{Local Force Sensor}}},
  author = {Balland, Yann and Absil, Luc and Pereira Dos Santos, Franck},
  year = 2024,
  month = sep,
  journal = {Physical Review Letters},
  volume = {133},
  number = {11},
  pages = {113403},
  issn = {0031-9007, 1079-7114},
  doi = {10.1103/PhysRevLett.133.113403},
  urldate = {2024-12-24},
  langid = {english}
}

@article{barrySensitivityOptimizationNVdiamond2020,
  title = {Sensitivity Optimization for {{NV-diamond}} Magnetometry},
  author = {Barry, John F. and Schloss, Jennifer M. and Bauch, Erik and Turner, Matthew J. and Hart, Connor A. and Pham, Linh M. and Walsworth, Ronald L.},
  year = 2020,
  month = mar,
  journal = {Reviews of Modern Physics},
  volume = {92},
  number = {1},
  pages = {015004},
  issn = {0034-6861, 1539-0756},
  doi = {10.1103/RevModPhys.92.015004},
  urldate = {2026-08-05},
  langid = {english}
}

@article{bayeExactNonrelativisticPolarizabilities2012,
  title = {Exact Nonrelativistic Polarizabilities of the Hydrogen Atom with the {{Lagrange-mesh}} Method},
  author = {Baye, Daniel},
  year = 2012,
  month = dec,
  journal = {Physical Review A},
  volume = {86},
  number = {6},
  pages = {062514},
  issn = {1050-2947, 1094-1622},
  doi = {10.1103/PhysRevA.86.062514},
  urldate = {2026-07-17},
  copyright = {http://link.aps.org/licenses/aps-default-license},
  langid = {english}
}

@article{bishtPrecisionTimekeepingAtomic2026,
  title = {Precision Timekeeping with Atomic Clocks: Evolution and Future Directions},
  shorttitle = {Precision Timekeeping with Atomic Clocks},
  author = {Bisht, Anjali and Arora, Poonam and Achanta, Venu Gopal},
  year = 2026,
  month = jun,
  journal = {Nature Communications},
  volume = {17},
  number = {1},
  pages = {5284},
  issn = {2041-1723},
  doi = {10.1038/s41467-026-73441-1},
  urldate = {2026-08-05},
  langid = {english}
}

@book{boninElectricDipolePolarizabilitiesAtoms1997,
  title = {Electric-{{Dipole Polarizabilities}} of {{Atoms}}, {{Molecules}}, and {{Clusters}}},
  author = {Bonin, Keith D and Kresin, Vitaly V},
  year = 1997,
  month = oct,
  publisher = {WORLD SCIENTIFIC},
  doi = {10.1142/2962},
  urldate = {2026-08-07},
  isbn = {978-981-02-2493-6 978-981-4261-27-2},
  langid = {english}
}

@article{bouchiatLinearStarkShifts2008,
  title = {Linear {{Stark Shifts}} to {{Measure}} the {{Fr Weak Nuclear Charge}} with {{Small Atom Samples}}},
  author = {Bouchiat, Marie-Anne},
  year = 2008,
  month = mar,
  journal = {Physical Review Letters},
  volume = {100},
  number = {12},
  eprint = {0711.0337},
  primaryclass = {physics.atom-ph},
  pages = {123003},
  issn = {0031-9007, 1079-7114},
  doi = {10.1103/PhysRevLett.100.123003},
  urldate = {2026-07-17},
  archiveprefix = {arXiv},
  langid = {english}
}

@article{bouchiatMagneticDipoleElectric1988,
  title = {Magnetic Dipole and Electric Quadrupole Amplitudes Induced by the Hyperfine Interaction in the {{Cesium 6S-7S}} Transition and the Parity Violation Calibration},
  author = {Bouchiat, C. and Piketty, C.A.},
  year = 1988,
  journal = {Journal de Physique},
  volume = {49},
  number = {11},
  pages = {1851--1856},
  issn = {0302-0738},
  doi = {10.1051/jphys:0198800490110185100},
  urldate = {2026-07-17},
  langid = {english}
}

@article{brueMagneticallyTunableFeshbach2012,
  title = {Magnetically {{Tunable Feshbach Resonances}} in {{Ultracold Li-Yb Mixtures}}},
  author = {Brue, Daniel A. and Hutson, Jeremy M.},
  year = 2012,
  month = jan,
  journal = {Physical Review Letters},
  volume = {108},
  number = {4},
  pages = {043201},
  issn = {0031-9007, 1079-7114},
  doi = {10.1103/PhysRevLett.108.043201},
  urldate = {2026-07-17},
  copyright = {http://link.aps.org/licenses/aps-default-license},
  langid = {english}
}

@article{bullisRamseySpectroscopy22023,
  title = {Ramsey {{Spectroscopy}} of the 2 {{S}} 1 / 2 {{Hyperfine Interval}} in {{Atomic Hydrogen}}},
  author = {Bullis, R. G. and Rasor, C. and Tavis, W. L. and Johnson, S. A. and Weiss, M. R. and Yost, D. C.},
  year = 2023,
  month = may,
  journal = {Physical Review Letters},
  volume = {130},
  number = {20},
  pages = {203001},
  issn = {0031-9007, 1079-7114},
  doi = {10.1103/PhysRevLett.130.203001},
  urldate = {2026-07-17},
  langid = {english}
}

@article{bungeRoothaanHartreeFockGroundStateAtomic1993,
  title = {Roothaan-{{Hartree-Fock Ground-State Atomic Wave Functions}}: {{Slater-Type Orbital Expansions}} and {{Expectation Values}} for {{Z}} = 2-54},
  shorttitle = {Roothaan-{{Hartree-Fock Ground-State Atomic Wave Functions}}},
  author = {Bunge, C.F. and Barrientos, J.A. and Bunge, A.V.},
  year = 1993,
  month = jan,
  journal = {Atomic Data and Nuclear Data Tables},
  volume = {53},
  number = {1},
  pages = {113--162},
  issn = {0092640X},
  doi = {10.1006/adnd.1993.1003},
  urldate = {2026-07-17},
  copyright = {https://www.elsevier.com/tdm/userlicense/1.0/},
  langid = {english}
}

@article{camparoSemiempiricalTheoryCarver2007,
  title = {Semiempirical Theory of {{Carver}} Rates in Alkali/Noble-Gas Systems},
  author = {Camparo, J. C.},
  year = 2007,
  month = jun,
  journal = {The Journal of Chemical Physics},
  volume = {126},
  number = {24},
  pages = {244310},
  issn = {0021-9606, 1089-7690},
  doi = {10.1063/1.2743958},
  urldate = {2026-07-17},
  langid = {english}
}

@article{casimirInfluenceRetardationLondonvan1948,
  title = {The {{Influence}} of {{Retardation}} on the {{London-van}} Der {{Waals Forces}}},
  author = {Casimir, H. B. G. and Polder, D.},
  year = 1948,
  month = feb,
  journal = {Physical Review},
  volume = {73},
  number = {4},
  pages = {360--372},
  issn = {0031-899X},
  doi = {10.1103/PhysRev.73.360},
  urldate = {2026-07-20},
  copyright = {http://link.aps.org/licenses/aps-default-license},
  langid = {english}
}

@article{chanTailoringOpticalMetamaterials2018,
  title = {Tailoring Optical Metamaterials to Tune the Atom-Surface {{Casimir-Polder}} Interaction},
  author = {Chan, Eng Aik and Aljunid, Syed Abdullah and Adamo, Giorgio and Laliotis, Athanasios and Ducloy, Martial and Wilkowski, David},
  year = 2018,
  month = feb,
  journal = {Science Advances},
  volume = {4},
  number = {2},
  pages = {eaao4223},
  issn = {2375-2548},
  doi = {10.1126/sciadv.aao4223},
  urldate = {2026-08-04},
  langid = {english}
}

@article{Chin2010,
  title = {Feshbach Resonances in Ultracold Gases},
  author = {Chin, Cheng and Julienne, Paul and Tiesinga, Eite},
  year = 2010,
  month = apr,
  journal = {Reviews of Modern Physics},
  volume = {82},
  number = {2},
  pages = {1225--1286},
  issn = {0034-6861},
  doi = {10.1103/RevModPhys.82.1225},
  urldate = {2011-07-04}
}

@book{cohen-tannoudjiQuantumMechanics22005,
  title = {Quantum {{Mechanics}}, {{Volume}} 2: {{Angular Momentum}}, {{Spin}}, and {{Approximation Methods}}},
  author = {{Cohen-Tannoudji}, Claude and Diu, Bernard and Lalo{\"e}, Franck},
  year = 2005,
  series = {Textbook Physics},
  edition = {Second},
  volume = {2},
  publisher = {Wiley},
  address = {New York},
  isbn = {978-0-471-16435-7},
  langid = {english}
}

@article{congSpindependentExoticInteractions2025,
  title = {Spin-Dependent Exotic Interactions},
  author = {Cong, Lei and Ji, Wei and Fadeev, Pavel and Ficek, Filip and Jiang, Min and Flambaum, Victor V. and Guan, Haosen and Jackson Kimball, Derek F. and Kozlov, Mikhail G. and Stadnik, Yevgeny V. and Budker, Dmitry},
  year = 2025,
  month = jun,
  journal = {Reviews of Modern Physics},
  volume = {97},
  number = {2},
  pages = {025005},
  issn = {0034-6861, 1539-0756},
  doi = {10.1103/RevModPhys.97.025005},
  urldate = {2026-08-05},
  langid = {english}
}

@article{darosColdAtomsMicromachined2020,
  title = {Cold Atoms in Micromachined Waveguides: {{A}} New Platform for Atom-Photon Interactions},
  shorttitle = {Cold Atoms in Micromachined Waveguides},
  author = {Da Ros, E. and Cooper, N. and Nute, J. and Hackermueller, L.},
  year = 2020,
  month = jul,
  journal = {Physical Review Research},
  volume = {2},
  number = {3},
  pages = {033098},
  issn = {2643-1564},
  doi = {10.1103/PhysRevResearch.2.033098},
  urldate = {2026-06-29},
  langid = {english}
}

@article{debordHollowCoreFiberTechnology2019,
  title = {Hollow-{{Core Fiber Technology}}: {{The Rising}} of ``{{Gas Photonics}}''},
  shorttitle = {Hollow-{{Core Fiber Technology}}},
  author = {Debord, Beno{\^i}t and Amrani, Foued and Vincetti, Luca and G{\'e}r{\^o}me, Fr{\'e}d{\'e}ric and Benabid, Fetah},
  year = 2019,
  month = feb,
  journal = {Fibers},
  volume = {7},
  number = {2},
  pages = {16},
  issn = {2079-6439},
  doi = {10.3390/fib7020016},
  urldate = {2026-06-29},
  langid = {english}
}

@article{dereviankoElectricDipolePolarizabilities2010a,
  title = {Electric Dipole Polarizabilities at Imaginary Frequencies for Hydrogen, the Alkali--Metal, Alkaline--Earth, and Noble Gas Atoms},
  author = {Derevianko, Andrei and Porsev, Sergey G. and Babb, James F.},
  year = 2010,
  month = may,
  journal = {Atomic Data and Nuclear Data Tables},
  volume = {96},
  number = {3},
  pages = {323--331},
  issn = {0092640X},
  doi = {10.1016/j.adt.2009.12.002},
  urldate = {2026-07-17},
  copyright = {https://www.elsevier.com/tdm/userlicense/1.0/},
  langid = {english}
}

@article{duSinglemoleculeScaleMagnetic2024,
  title = {Single-Molecule Scale Magnetic Resonance Spectroscopy Using Quantum Diamond Sensors},
  author = {Du, Jiangfeng and Shi, Fazhan and Kong, Xi and Jelezko, Fedor and Wrachtrup, J{\"o}rg},
  year = 2024,
  month = may,
  journal = {Reviews of Modern Physics},
  volume = {96},
  number = {2},
  pages = {025001},
  issn = {0034-6861, 1539-0756},
  doi = {10.1103/RevModPhys.96.025001},
  urldate = {2026-08-05},
  langid = {english}
}

@article{duttaEffectsHigherorderCasimirPolder2024,
  title = {Effects of Higher-Order {{Casimir-Polder}} Interactions on {{Rydberg}} Atom Spectroscopy},
  author = {Dutta, B. and Carvalho, J. C. De Aquino and {Garcia-Arellano}, G. and Pedri, P. and Laliotis, A. and Boldt, C. and Kaushal, J. and Scheel, S.},
  year = 2024,
  month = may,
  journal = {Physical Review Research},
  volume = {6},
  number = {2},
  pages = {L022035},
  issn = {2643-1564},
  doi = {10.1103/PhysRevResearch.6.L022035},
  urldate = {2026-08-04},
  langid = {english}
}

@phdthesis{duttaProbingCesiumRydberg,
  title = {Probing Cesium {{Rydberg}} Atoms Close to a Sapphire Surface},
  author = {Dutta, Biplab},
  year = 2024,
  month = may,
  langid = {english},
  school = {Sorbonne}
}

@article{duttaSelectiveReflectionCasimirPolder2025,
  title = {Selective Reflection {{Casimir-Polder}} Spectroscopy in Vapor Cells: {{The}} Influence of the Thermal Velocity Distribution},
  shorttitle = {Selective Reflection {{Casimir-Polder}} Spectroscopy in Vapor Cells},
  author = {Dutta, B. and Boldt, C. and {Garcia-Arellano}, G. and Ducloy, M. and Pedri, P. and Scheel, S. and Laliotis, A.},
  year = 2025,
  month = jul,
  journal = {Physical Review A},
  volume = {112},
  number = {1},
  pages = {012815},
  issn = {2469-9926, 2469-9934},
  doi = {10.1103/pygs-cvr9},
  urldate = {2026-06-29},
  langid = {english}
}

@article{dzubaOffdiagonalHyperfineInteraction2000,
  title = {Off-Diagonal Hyperfine Interaction and Parity Nonconservation in Cesium},
  author = {Dzuba, V. A. and Flambaum, V. V.},
  year = 2000,
  month = oct,
  journal = {Physical Review A},
  volume = {62},
  number = {5},
  pages = {052101},
  issn = {1050-2947, 1094-1622},
  doi = {10.1103/PhysRevA.62.052101},
  urldate = {2026-07-17},
  copyright = {http://link.aps.org/licenses/aps-default-license},
  langid = {english}
}

@article{eshelRoleAdiabaticityControlling2017,
  title = {Role of Adiabaticity in Controlling Alkali-Metal Fine-Structure Mixing Induced by Rare Gases},
  author = {Eshel, Ben and Cardoza, Joseph A. and Weeks, David E. and Perram, Glen P.},
  year = 2017,
  month = apr,
  journal = {Physical Review A},
  volume = {95},
  number = {4},
  pages = {042708},
  issn = {2469-9926, 2469-9934},
  doi = {10.1103/PhysRevA.95.042708},
  urldate = {2026-07-15},
  copyright = {http://link.aps.org/licenses/aps-default-license},
  langid = {english}
}

@article{fichetVanWaalsInteractions1995,
  title = {Van Der {{Waals}} Interactions between Excited-State Atoms and Dispersive Dielectric Surfaces},
  author = {Fichet, M. and Schuller, F. and Bloch, D. and Ducloy, M.},
  year = 1995,
  month = feb,
  journal = {Physical Review A},
  volume = {51},
  number = {2},
  pages = {1553--1564},
  issn = {1050-2947, 1094-1622},
  doi = {10.1103/PhysRevA.51.1553},
  urldate = {2026-08-12},
  copyright = {http://link.aps.org/licenses/aps-default-license},
  langid = {english}
}

@article{fleigVariationalCalculationHyperfine2025,
  title = {Variational Calculation of the Hyperfine {{Stark}} Effect in Atomic {{Rb}} 87 , {{Cs}} 133 , and {{Tm}} 169},
  author = {Fleig, Timo},
  year = 2025,
  month = nov,
  journal = {Physical Review A},
  volume = {112},
  number = {5},
  pages = {052802},
  issn = {2469-9926, 2469-9934},
  doi = {10.1103/cy87-s81k},
  urldate = {2026-07-17},
  langid = {english}
}

@article{fortaghMagneticMicrotrapsUltracold2007,
  title = {Magnetic Microtraps for Ultracold Atoms},
  author = {Fort{\'a}gh, J{\'o}zsef and Zimmermann, Claus},
  year = 2007,
  month = feb,
  journal = {Reviews of Modern Physics},
  volume = {79},
  number = {1},
  pages = {235--289},
  issn = {0034-6861, 1539-0756},
  doi = {10.1103/RevModPhys.79.235},
  urldate = {2026-08-05},
  copyright = {http://link.aps.org/licenses/aps-default-license},
  langid = {english}
}

@article{garcia-vidalSpoofSurfacePlasmon2022,
  title = {Spoof Surface Plasmon Photonics},
  author = {{Garcia-Vidal}, Francisco J. and {Fern{\'a}ndez-Dom{\'i}nguez}, Antonio I. and {Martin-Moreno}, Luis and Zhang, Hao Chi and Tang, Wenxuan and Peng, Ruwen and Cui, Tie Jun},
  year = 2022,
  month = may,
  journal = {Reviews of Modern Physics},
  volume = {94},
  number = {2},
  pages = {025004},
  issn = {0034-6861, 1539-0756},
  doi = {10.1103/RevModPhys.94.025004},
  urldate = {2026-08-28},
  langid = {english}
}

@article{gongNonlinearPressureShifts2008,
  title = {Nonlinear {{Pressure Shifts}} of {{Alkali-Metal Atoms}} in {{Inert Gases}}},
  author = {Gong, F. and Jau, Y.-Y. and Happer, W.},
  year = 2008,
  month = jun,
  journal = {Physical Review Letters},
  volume = {100},
  number = {23},
  pages = {233002},
  issn = {0031-9007, 1079-7114},
  doi = {10.1103/PhysRevLett.100.233002},
  urldate = {2026-07-17},
  copyright = {http://link.aps.org/licenses/aps-default-license},
  langid = {english}
}

@book{gradstejnTableIntegralsSeries2009,
  title = {Table of Integrals, Series and Products},
  author = {Grad{\v s}tejn, Izrail' S. and Ry{\v z}ik, Josif M. and Jeffrey, Alan and Zwillinger, Daniel and Grad{\v s}tejn, Izrail' S.},
  year = 2009,
  edition = {7. ed., [3. Nachdr.]},
  publisher = {Elsevier Acad. Press},
  address = {Amsterdam},
  isbn = {978-0-12-373637-6},
  langid = {english}
}

@book{griffithsIntroductionQuantumMechanics2018,
  title = {Introduction to Quantum Mechanics},
  author = {Griffiths, David J. and Schroeter, Darrell F.},
  year = 2018,
  edition = {Third edition},
  publisher = {Cambridge University Press},
  address = {Cambridge},
  doi = {10.1017/9781316995433},
  isbn = {978-1-107-18963-8 978-1-316-99543-3},
  langid = {english}
}

@article{grossQuantumGasMicroscopy2021,
  title = {Quantum Gas Microscopy for Single Atom and Spin Detection},
  author = {Gross, Christian and Bakr, Waseem S.},
  year = 2021,
  month = dec,
  journal = {Nature Physics},
  volume = {17},
  number = {12},
  pages = {1316--1323},
  issn = {1745-2473, 1745-2481},
  doi = {10.1038/s41567-021-01370-5},
  urldate = {2022-02-02},
  langid = {english}
}

@article{hansonSpinsFewelectronQuantum2007,
  title = {Spins in Few-Electron Quantum Dots},
  author = {Hanson, R. and Kouwenhoven, L. P. and Petta, J. R. and Tarucha, S. and Vandersypen, L. M. K.},
  year = 2007,
  month = oct,
  journal = {Reviews of Modern Physics},
  volume = {79},
  number = {4},
  pages = {1217--1265},
  issn = {0034-6861, 1539-0756},
  doi = {10.1103/RevModPhys.79.1217},
  urldate = {2026-08-05},
  copyright = {http://link.aps.org/licenses/aps-default-license},
  langid = {english}
}

@book{happerOpticallyPumpedAtoms2010,
  title = {Optically {{Pumped Atoms}}},
  author = {Happer, William},
  year = 2010,
  edition = {1st ed},
  publisher = {John Wiley \& Sons, Incorporated},
  address = {Weinheim},
  collaborator = {Jau, Yuan-Yu and Walker, Thad},
  isbn = {978-3-527-40707-1 978-3-527-62951-0},
  langid = {english}
}

@article{hermanFrequencyShiftsHyperfine1961,
  title = {Frequency {{Shifts}} in {{Hyperfine Splitting}} of {{Alkalis}}: A {{Correction}}},
  shorttitle = {Frequency {{Shifts}} in {{Hyperfine Splitting}} of {{Alkalis}}},
  author = {Herman, R. and Margenau, H.},
  year = 1961,
  month = may,
  journal = {Physical Review},
  volume = {122},
  number = {4},
  pages = {1204--1206},
  issn = {0031-899X},
  doi = {10.1103/PhysRev.122.1204},
  urldate = {2026-07-14},
  copyright = {http://link.aps.org/licenses/aps-default-license},
  langid = {english}
}

@article{hindmarshCollisionBroadeningSpectral1973,
  title = {Collision Broadening of Spectral Lines by Neutral Atoms},
  author = {Hindmarsh, W.R. and Farr, Judith M.},
  year = 1973,
  journal = {Progress in Quantum Electronics},
  volume = {2},
  pages = {141--214},
  issn = {00796727},
  doi = {10.1016/0079-6727(73)90005-0},
  urldate = {2026-07-17},
  langid = {english}
}

@article{ishikawaCollisionalShiftsHyperfine2024,
  title = {Collisional Shifts of Hyperfine Resonances of Ground-State Atoms Calculated with Modified Pseudopotential and Orthogonalization},
  author = {Ishikawa, Kiyoshi},
  year = 2024,
  month = may,
  journal = {Applied Physics B},
  volume = {130},
  number = {5},
  pages = {71},
  issn = {0946-2171, 1432-0649},
  doi = {10.1007/s00340-024-08203-2},
  urldate = {2026-07-17},
  langid = {english}
}

@article{ishikawaFlyingCharacterizationColliding2023,
  title = {Flying Characterization of Colliding Partners by Hyperfine Splitting Frequency of Neutral Paramagnetic Atoms},
  author = {Ishikawa, Kiyoshi},
  year = 2023,
  month = aug,
  journal = {The Journal of Chemical Physics},
  volume = {159},
  number = {6},
  publisher = {AIP Publishing},
  issn = {0021-9606, 1089-7690},
  doi = {10.1063/5.0161491},
  urldate = {2025-07-14},
  langid = {english}
}

@article{ishikawaNoblegasAtomsCharacterized2022,
  title = {Noble-Gas Atoms Characterized by Hyperfine Frequency Shift of Lithium Atom},
  author = {Ishikawa, Kiyoshi},
  year = 2022,
  month = apr,
  journal = {The Journal of Chemical Physics},
  volume = {156},
  number = {14},
  pages = {144301},
  issn = {0021-9606, 1089-7690},
  doi = {10.1063/5.0085859},
  urldate = {2026-07-17},
  langid = {english}
}

@article{ishikawaPseudopotentialAnalysisHyperfine2023,
  title = {Pseudopotential Analysis on Hyperfine Splitting Frequency Shift of Alkali-Metal Atoms in Noble Gases, Revisited},
  author = {Ishikawa, Kiyoshi},
  year = 2023,
  month = feb,
  journal = {The Journal of Chemical Physics},
  volume = {158},
  number = {8},
  pages = {084306},
  issn = {0021-9606, 1089-7690},
  doi = {10.1063/5.0138434},
  urldate = {2026-07-17},
  langid = {english}
}

@article{itanoShift121982,
  title = {Shift of {{S}} 1 2 2 Hyperfine Splittings Due to Blackbody Radiation},
  author = {Itano, Wayne M. and Lewis, L. L. and Wineland, D. J.},
  year = 1982,
  month = feb,
  journal = {Physical Review A},
  volume = {25},
  number = {2},
  pages = {1233--1235},
  issn = {0556-2791},
  doi = {10.1103/PhysRevA.25.1233},
  urldate = {2026-07-17},
  copyright = {http://link.aps.org/licenses/aps-default-license},
  langid = {english}
}

@article{jentschuraQuantumElectrodynamicCorrections2006,
  title = {Quantum Electrodynamic Corrections to the Hyperfine Structure of Excited {{S}} States},
  author = {Jentschura, Ulrich D. and Yerokhin, Vladimir A.},
  year = 2006,
  month = jun,
  journal = {Physical Review A},
  volume = {73},
  number = {6},
  pages = {062503},
  issn = {1050-2947, 1094-1622},
  doi = {10.1103/PhysRevA.73.062503},
  urldate = {2026-07-17},
  copyright = {http://link.aps.org/licenses/aps-default-license},
  langid = {english}
}

@article{jones_ultracold_2006,
  title = {Ultracold Photoassociation Spectroscopy: {{Long-range}} Molecules and Atomic Scattering},
  author = {Jones, Kevin M. and Tiesinga, Eite and Lett, Paul D. and Julienne, Paul S.},
  year = 2006,
  journal = {Reviews of Modern Physics},
  volume = {78},
  number = {2},
  pages = {483--535},
  issn = {00346861},
  doi = {10.1103/RevModPhys.78.483},
  isbn = {0034-6861 1539-0756}
}

@article{karshenboimHyperfineStructureHydrogen2002,
  title = {Hyperfine Structure in Hydrogen and Helium Ion},
  author = {Karshenboim, Savely G. and Ivanov, Vladimir G.},
  year = 2002,
  month = jan,
  journal = {Physics Letters B},
  volume = {524},
  number = {3-4},
  pages = {259--264},
  issn = {03702693},
  doi = {10.1016/S0370-2693(01)01394-6},
  urldate = {2026-07-17},
  copyright = {https://www.elsevier.com/tdm/userlicense/1.0/},
  langid = {english}
}

@article{karshenboimPossibilitiesLaboratorySearches,
  title = {Some Possibilities for Laboratory Searches for Variations of Fundamental Constants},
  author = {Karshenboim, Savely G},
  year = 2000,
  journal = {Canadian Journal of Physics},
  volume = {78},
  pages = {639--678},
  langid = {english}
}

@article{klimchitskayaCasimirForceReal2009,
  title = {The {{Casimir}} Force between Real Materials: {{Experiment}} and Theory},
  shorttitle = {The {{Casimir}} Force between Real Materials},
  author = {Klimchitskaya, G. L. and Mohideen, U. and Mostepanenko, V. M.},
  year = 2009,
  month = dec,
  journal = {Reviews of Modern Physics},
  volume = {81},
  number = {4},
  pages = {1827--1885},
  issn = {0034-6861, 1539-0756},
  doi = {10.1103/RevModPhys.81.1827},
  urldate = {2026-08-05},
  copyright = {http://link.aps.org/licenses/aps-default-license},
  langid = {english}
}

@article{laliotisAtomsurfacePhysicsReview2021a,
  title = {Atom-Surface Physics: {{A}} Review},
  shorttitle = {Atom-Surface Physics},
  author = {Laliotis, Athanasios and Lu, Bing-Sui and Ducloy, Martial and Wilkowski, David},
  year = 2021,
  month = dec,
  journal = {AVS Quantum Science},
  volume = {3},
  number = {4},
  pages = {043501},
  issn = {2639-0213},
  doi = {10.1116/5.0063701},
  urldate = {2026-08-05},
  langid = {english}
}

@article{lecoffreMeasurementCasimirPolderInteraction2025,
  title = {Measurement of {{Casimir-Polder}} Interaction for Slow Atoms through a Material Grating},
  author = {Lecoffre, Julien and Hadi, Ayoub and Bruneau, Matthieu and Garcion, Charles and Fabre, Nathalie and Charron, {\'E}ric and Gaaloul, Naceur and Dutier, Gabriel and Bouton, Quentin},
  year = 2025,
  month = mar,
  journal = {Physical Review Research},
  volume = {7},
  number = {1},
  pages = {013232},
  issn = {2643-1564},
  doi = {10.1103/PhysRevResearch.7.013232},
  urldate = {2026-06-29},
  langid = {english}
}

@article{lennard-jonesProcessesAdsorptionDiffusion1932,
  title = {Processes of Adsorption and Diffusion on Solid Surfaces},
  author = {{Lennard-Jones}, J. E.},
  year = 1932,
  journal = {Transactions of the Faraday Society},
  volume = {28},
  pages = {333},
  issn = {0014-7672},
  doi = {10.1039/tf9322800333},
  urldate = {2026-08-06},
  langid = {english}
}

@article{ludlowOpticalAtomicClocks2015,
  title = {Optical Atomic Clocks},
  author = {Ludlow, Andrew D. and Boyd, Martin M. and Ye, Jun and Peik, E. and Schmidt, P. O.},
  year = 2015,
  month = jun,
  journal = {Reviews of Modern Physics},
  volume = {87},
  number = {2},
  pages = {637--701},
  issn = {0034-6861, 1539-0756},
  doi = {10.1103/RevModPhys.87.637},
  urldate = {2026-08-13},
  copyright = {http://link.aps.org/licenses/aps-default-license},
  langid = {english}
}

@article{margenauPressureEffectsSpectral1936,
  title = {Pressure {{Effects}} on {{Spectral Lines}}},
  author = {Margenau, Henry and Watson, William W.},
  year = 1936,
  month = jan,
  journal = {Reviews of Modern Physics},
  volume = {8},
  number = {1},
  pages = {22--53},
  issn = {0034-6861},
  doi = {10.1103/RevModPhys.8.22},
  urldate = {2026-07-17},
  copyright = {http://link.aps.org/licenses/aps-default-license},
  langid = {english}
}

@phdthesis{mcguyerAtomicPhysicsVaporcell,
  title = {Atomic Physics with Vapor-Cell Clocks},
  author = {McGuyer, Bart Hunter},
  year = 2012,
  langid = {english},
  school = {Princeton University}
}

@article{mcguyerHyperfineFrequencies872011,
  title = {Hyperfine Frequencies of 87 {{Rb}} and 133 {{Cs}} Atoms in {{Xe}} Gas},
  author = {McGuyer, B. H. and Xia, T. and Jau, Y.-Y. and Happer, W.},
  year = 2011,
  month = sep,
  journal = {Physical Review A},
  volume = {84},
  number = {3},
  pages = {030501},
  issn = {1050-2947, 1094-1622},
  doi = {10.1103/PhysRevA.84.030501},
  urldate = {2026-07-17},
  copyright = {http://link.aps.org/licenses/aps-default-license},
  langid = {english}
}

@article{mcguyerHyperfinefrequencyShiftsAlkalimetal2013,
  title = {Hyperfine-Frequency Shifts of Alkali-Metal Atoms during Long-Range Collisions},
  author = {McGuyer, B. H.},
  year = 2013,
  month = may,
  journal = {Physical Review A},
  volume = {87},
  number = {5},
  publisher = {American Physical Society (APS)},
  issn = {1050-2947, 1094-1622},
  doi = {10.1103/physreva.87.054702},
  urldate = {2025-07-14},
  copyright = {http://link.aps.org/licenses/aps-default-license},
  langid = {english}
}

@article{mcguyerIsotopeStudyNonlinear2023a,
  title = {Isotope Study of the Nonlinear Pressure Shifts of {{85Rb}} and {{87Rb}} Hyperfine Resonances in {{Ar}}, {{Kr}}, and {{Xe}} Buffer Gases},
  author = {McGuyer, B. H.},
  year = 2023,
  month = apr,
  journal = {The Journal of Chemical Physics},
  volume = {158},
  number = {14},
  pages = {144304},
  issn = {0021-9606, 1089-7690},
  doi = {10.1063/5.0145919},
  urldate = {2026-07-17},
  langid = {english}
}

@article{mcleanROOTHAANHARTREEFOCKATOMICWAVE,
  title = {Roothaan-{{Hartree-Fock Aomic Wave Functions Slater Basis-Set Expansions}} for {{Z}} = 55-92},
  author = {Mclean, A D and Mclean, R. S.},
  year = 1981,
  journal = {Atomic Data and Nuclear Data Tables},
  volume = {26},
  pages = {197--381},
  langid = {english}
}

@article{mitroyTheoryApplicationsAtomic2010a,
  title = {Theory and Applications of Atomic and Ionic Polarizabilities},
  author = {Mitroy, J and Safronova, M S and Clark, Charles W},
  year = 2010,
  month = oct,
  journal = {Journal of Physics B: Atomic, Molecular and Optical Physics},
  volume = {43},
  number = {20},
  pages = {202001},
  issn = {0953-4075, 1361-6455},
  doi = {10.1088/0953-4075/43/20/202001},
  urldate = {2026-07-17},
  langid = {english}
}

@book{nationalinstituteofstandardsandtechnologyNISTHandbookMathematical2010,
  title = {{{NIST}} Handbook of Mathematical Functions},
  author = {{National Institute of Standards and Technology}},
  editor = {Olver, Frank W. J. and Lozier, Daniel W. and Boisvert, Ronald F. and Clark, Charles W.},
  year = 2010,
  publisher = {Cambridge University Press},
  address = {Cambridge New York Melbourne},
  isbn = {978-0-521-19225-5 978-0-521-14063-8},
  langid = {english}
}

@article{norciaFrequencyMeasurementsSuperradiance2018,
  title = {Frequency {{Measurements}} of {{Superradiance}} from the {{Strontium Clock Transition}}},
  author = {Norcia, Matthew A. and Cline, Julia R. K. and Muniz, Juan A. and Robinson, John M. and Hutson, Ross B. and Goban, Akihisa and Marti, G. Edward and Ye, Jun and Thompson, James K.},
  year = 2018,
  month = may,
  journal = {Physical Review X},
  volume = {8},
  number = {2},
  pages = {021036},
  issn = {2160-3308},
  doi = {10.1103/PhysRevX.8.021036},
  urldate = {2026-08-28},
  langid = {english}
}

@article{okabaLambDickeSpectroscopyAtoms2014,
  title = {Lamb-{{Dicke}} Spectroscopy of Atoms in a Hollow-Core Photonic Crystal Fibre},
  author = {Okaba, Shoichi and Takano, Tetsushi and Benabid, Fetah and Bradley, Tom and Vincetti, Luca and Maizelis, Zakhar and Yampol'skii, Valery and Nori, Franco and Katori, Hidetoshi},
  year = 2014,
  month = jun,
  journal = {Nature Communications},
  volume = {5},
  number = {1},
  pages = {4096},
  issn = {2041-1723},
  doi = {10.1038/ncomms5096},
  urldate = {2026-06-29},
  langid = {english}
}

@article{oretoBuffergasinducedShiftBroadening2004,
  title = {Buffer-Gas-Induced Shift and Broadening of Hyperfine Resonances in Alkali-Metal Vapors},
  author = {Oreto, P. J. and Jau, Y.-Y. and Post, A. B. and Kuzma, N. N. and Happer, W.},
  year = 2004,
  month = apr,
  journal = {Physical Review A},
  volume = {69},
  number = {4},
  pages = {042716},
  issn = {1050-2947, 1094-1622},
  doi = {10.1103/PhysRevA.69.042716},
  urldate = {2026-07-17},
  copyright = {http://link.aps.org/licenses/aps-default-license},
  langid = {english}
}

@article{perssonTableHyperfineAnomaly2023,
  title = {Table of Hyperfine Anomaly in Atomic Systems --- 2023},
  author = {Persson, J.R.},
  year = 2023,
  month = nov,
  journal = {Atomic Data and Nuclear Data Tables},
  volume = {154},
  pages = {101589},
  issn = {0092640X},
  doi = {10.1016/j.adt.2023.101589},
  urldate = {2026-07-17},
  langid = {english}
}

@article{pezzeQuantumMetrologyNonclassical2018,
  title = {Quantum Metrology with Nonclassical States of Atomic Ensembles},
  author = {Pezz{\`e}, Luca and Smerzi, Augusto and Oberthaler, Markus K. and Schmied, Roman and Treutlein, Philipp},
  year = 2018,
  month = sep,
  journal = {Reviews of Modern Physics},
  volume = {90},
  number = {3},
  pages = {035005},
  issn = {0034-6861, 1539-0756},
  doi = {10.1103/RevModPhys.90.035005},
  urldate = {2026-08-05},
  langid = {english}
}

@article{quintStringentConstraintsNew2026,
  title = {Stringent {{Constraints}} on {{New Pseudoscalar}} and {{Vector Bosons}} from {{Precision Hyperfine Splitting Measurements}}},
  author = {Quint, Cedric and Hei{\ss}e, Fabian and Jaeckel, Joerg and Leimenstoll, Lutz and Keitel, Christoph H. and Harman, Zolt{\'a}n},
  year = 2026,
  month = mar,
  journal = {Physical Review Letters},
  volume = {136},
  number = {11},
  pages = {113001},
  issn = {0031-9007, 1079-7114},
  doi = {10.1103/rvt1-93v2},
  urldate = {2026-04-23},
  langid = {english}
}

@article{ritterCouplingThermalAtomic2018,
  title = {Coupling {{Thermal Atomic Vapor}} to {{Slot Waveguides}}},
  author = {Ritter, Ralf and Gruhler, Nico and Dobbertin, Helge and K{\"u}bler, Harald and Scheel, Stefan and Pernice, Wolfram and Pfau, Tilman and L{\"o}w, Robert},
  year = 2018,
  month = may,
  journal = {Physical Review X},
  volume = {8},
  number = {2},
  pages = {021032},
  issn = {2160-3308},
  doi = {10.1103/PhysRevX.8.021032},
  urldate = {2026-08-11},
  langid = {english}
}

@article{safronovaRelativisticManybodyCalculations1999,
  title = {Relativistic Many-Body Calculations of Energy Levels, Hyperfine Constants, Electric-Dipole Matrix Elements, and Static Polarizabilities for Alkali-Metal Atoms},
  author = {Safronova, M. S. and Johnson, W. R. and Derevianko, A.},
  year = 1999,
  month = dec,
  journal = {Physical Review A},
  volume = {60},
  number = {6},
  pages = {4476--4487},
  issn = {1050-2947, 1094-1622},
  doi = {10.1103/PhysRevA.60.4476},
  urldate = {2026-07-17},
  copyright = {http://link.aps.org/licenses/aps-default-license},
  langid = {english}
}

@article{safronovaSearchNewPhysics2018,
  title = {Search for New Physics with Atoms and Molecules},
  author = {Safronova, M. S. and Budker, D. and DeMille, D. and Kimball, Derek F. Jackson and Derevianko, A. and Clark, Charles W.},
  year = 2018,
  month = jun,
  journal = {Reviews of Modern Physics},
  volume = {90},
  number = {2},
  pages = {025008},
  issn = {0034-6861, 1539-0756},
  doi = {10.1103/RevModPhys.90.025008},
  urldate = {2026-08-05},
  langid = {english}
}

@article{sanchezMatrixelementCalculationsHydrogenlike1992,
  title = {Matrix-Element Calculations for Hydrogenlike Atoms},
  author = {S{\'a}nchez, M. L. and Moreno, B. and L{\'o}pez Pi{\~n}eiro, A.},
  year = 1992,
  month = dec,
  journal = {Physical Review A},
  volume = {46},
  number = {11},
  pages = {6908--6913},
  issn = {1050-2947, 1094-1622},
  doi = {10.1103/PhysRevA.46.6908},
  urldate = {2026-07-17},
  copyright = {http://link.aps.org/licenses/aps-default-license},
  langid = {english}
}

@article{sargsyanCompetingVanWaals2023,
  title = {Competing van Der {{Waals}} and Dipole-Dipole Interactions in Optical Nanocells at Thicknesses below 100 Nm},
  author = {Sargsyan, Armen and Momier, Rodolphe and Leroy, Claude and Sarkisyan, David},
  year = 2023,
  month = sep,
  journal = {Physics Letters A},
  volume = {483},
  pages = {129069},
  issn = {03759601},
  doi = {10.1016/j.physleta.2023.129069},
  urldate = {2026-06-29},
  langid = {english}
}

@misc{sargsyanDopplerfreeSpectroscopyCs,
  title = {Doppler-Free Spectroscopy of the {{Cs 6S1}}/2 - {{7P3}}/2 Atomic Transition at 456 Nm in a Nanometric-Thick Vapor Layer},
  author = {Sargsyan, Armen and Klinger, Emmanuel and Boudot, Rodolphe and Sarkisyan, David},
  year = 2025,
  langid = {english}
}

@article{sargsyanFeaturesVanWaals2026,
  title = {Features of the van Der {{Waals Interaction}} on the {{Cesium}} 6 {{S}} 1 / 2 {$\rightarrow$} 7 {{P}} 3 / 2 Transition in an Optical Nanocell},
  author = {Sargsyan, Armen and Gogyan, Anahit and Sarkisyan, David},
  year = 2026,
  month = may,
  journal = {Spectrochimica Acta Part B: Atomic Spectroscopy},
  volume = {239},
  pages = {107493},
  issn = {05848547},
  doi = {10.1016/j.sab.2026.107493},
  urldate = {2026-07-15},
  langid = {english}
}

@article{sargsyanStudyInteractionRubidium2023,
  title = {Study of the {{Interaction}} of {{Rubidium Atoms}} with {{Sapphire Surface Using Spectroscopic Nanocells}}},
  author = {Sargsyan, A.},
  year = 2023,
  month = sep,
  journal = {Journal of Applied Spectroscopy},
  volume = {90},
  number = {4},
  pages = {731--735},
  issn = {0021-9037, 1573-8647},
  doi = {10.1007/s10812-023-01588-6},
  urldate = {2026-06-29},
  langid = {english}
}

@article{sargsyanStudyVanWaals2025,
  title = {Study of the {{Van}} Der {{Waals Effect}} in {{Potassium Atomic Vapours}}},
  author = {Sargsyan, A. D. and Sarkisyan, D. H.},
  year = 2025,
  month = sep,
  journal = {Journal of Contemporary Physics (Armenian Academy of Sciences)},
  volume = {60},
  number = {3},
  pages = {261--266},
  issn = {1068-3372, 1934-9378},
  doi = {10.1134/S1068337225700707},
  urldate = {2026-07-01},
  langid = {english}
}

@article{stuartDifferentialStarkShifts1980a,
  title = {Differential {{Stark}} Shifts in the Hydrogen Maser},
  author = {Stuart, James G. and Larson, Daniel J. and Ramsey, Norman F.},
  year = 1980,
  month = nov,
  journal = {Physical Review A},
  volume = {22},
  number = {5},
  pages = {2092--2097},
  issn = {0556-2791},
  doi = {10.1103/PhysRevA.22.2092},
  urldate = {2026-07-17},
  copyright = {http://link.aps.org/licenses/aps-default-license},
  langid = {english}
}

@book{vanierQuantumPhysicsAtomic1989,
  title = {The Quantum Physics of Atomic Frequency Standards},
  author = {Vanier, Jacques and Audoin, Claude},
  year = 1989,
  publisher = {Hilger},
  address = {Bristol},
  isbn = {978-0-85274-434-5},
  langid = {english}
}

@book{vanierQuantumPhysicsAtomic2025,
  title = {The Quantum Physics of Atomic Frequency Standards: Recent Developments},
  shorttitle = {The Quantum Physics of Atomic Frequency Standards},
  author = {Vanier, Jacques and Tomescu, Cipriana},
  year = 2025,
  edition = {Second edition},
  publisher = {CRC PRESS},
  address = {S.l.},
  doi = {10.1201/9781003440932},
  isbn = {978-1-032-56579-8 978-1-040-26919-0 978-1-003-44093-2 978-1-040-26917-6},
  langid = {english}
}

@article{whittakerSpectroscopicDetectionAtomsurface2015,
  title = {Spectroscopic Detection of Atom-Surface Interactions in an Atomic-Vapor Layer with Nanoscale Thickness},
  author = {Whittaker, K. A. and Keaveney, J. and Hughes, I. G. and Sargsyan, A. and Sarkisyan, D. and Adams, C. S.},
  year = 2015,
  month = nov,
  journal = {Physical Review A},
  volume = {92},
  number = {5},
  pages = {052706},
  issn = {1050-2947, 1094-1622},
  doi = {10.1103/PhysRevA.92.052706},
  urldate = {2026-07-17},
  copyright = {http://creativecommons.org/licenses/by/3.0/},
  langid = {english}
}

@article{wuWallInteractionsSpinpolarized2021,
  title = {Wall Interactions of Spin-Polarized Atoms},
  author = {Wu, Zhen},
  year = 2021,
  month = sep,
  journal = {Reviews of Modern Physics},
  volume = {93},
  number = {3},
  pages = {035006},
  issn = {0034-6861, 1539-0756},
  doi = {10.1103/RevModPhys.93.035006},
  urldate = {2026-07-17},
  langid = {english}
}

@article{wylieQuantumElectrodynamicsInterface1984,
  title = {Quantum Electrodynamics near an Interface},
  author = {Wylie, J. M. and Sipe, J. E.},
  year = 1984,
  month = sep,
  journal = {Physical Review A},
  volume = {30},
  number = {3},
  pages = {1185--1193},
  issn = {0556-2791},
  doi = {10.1103/PhysRevA.30.1185},
  urldate = {2026-08-28},
  copyright = {http://link.aps.org/licenses/aps-default-license},
  langid = {english}
}

@article{yangLaserSpectroscopyStudy2023,
  title = {Laser Spectroscopy for the Study of Exotic Nuclei},
  author = {Yang, X.F. and Wang, S.J. and Wilkins, S.G. and Ruiz, R.F. Garcia},
  year = 2023,
  month = mar,
  journal = {Progress in Particle and Nuclear Physics},
  volume = {129},
  pages = {104005},
  issn = {01466410},
  doi = {10.1016/j.ppnp.2022.104005},
  urldate = {2026-08-04},
  langid = {english}
}

@article{zhangLongrangeDispersionInteractions2007,
  title = {Long-Range Dispersion Interactions. {{I}}. {{Formalism}} for Two Heteronuclear Atoms},
  author = {Zhang, J.-Y. and Mitroy, J.},
  year = 2007,
  month = aug,
  journal = {Physical Review A},
  volume = {76},
  number = {2},
  pages = {022705},
  issn = {1050-2947, 1094-1622},
  doi = {10.1103/PhysRevA.76.022705},
  urldate = {2026-07-17},
  copyright = {http://link.aps.org/licenses/aps-default-license},
  langid = {english}
}

@article{zhangLongrangeDispersionInteractions2017a,
  title = {Long-Range Dispersion Interactions between Excited States of {{K}} and Rare-Gas Atoms},
  author = {Zhang, Deng-Hong and Xu, Ya-Bin and Jiang, Jun and Jiang, Li and Xie, Lu-You and Dong, Chen-Zhong},
  year = 2017,
  month = jul,
  journal = {Journal of Physics: Conference Series},
  volume = {875},
  pages = {082008},
  issn = {1742-6588, 1742-6596},
  doi = {10.1088/1742-6596/875/9/082008},
  urldate = {2026-07-15},
  copyright = {http://iopscience.iop.org/info/page/text-and-data-mining},
  langid = {english}
}

@article{zuchowskiUltracoldRbSrMolecules2010,
  title = {Ultracold {{RbSr Molecules Can Be Formed}} by {{Magnetoassociation}}},
  author = {{\.Z}uchowski, Piotr S. and Aldegunde, J. and Hutson, Jeremy M.},
  year = 2010,
  month = oct,
  journal = {Physical Review Letters},
  volume = {105},
  number = {15},
  pages = {153201},
  issn = {0031-9007, 1079-7114},
  doi = {10.1103/PhysRevLett.105.153201},
  urldate = {2026-07-17},
  copyright = {http://link.aps.org/licenses/aps-default-license},
  langid = {english}
}

@article{zwanenburgSiliconQuantumElectronics2013,
  title = {Silicon Quantum Electronics},
  author = {Zwanenburg, Floris A. and Dzurak, Andrew S. and Morello, Andrea and Simmons, Michelle Y. and Hollenberg, Lloyd C. L. and Klimeck, Gerhard and Rogge, Sven and Coppersmith, Susan N. and Eriksson, Mark A.},
  year = 2013,
  month = jul,
  journal = {Reviews of Modern Physics},
  volume = {85},
  number = {3},
  pages = {961--1019},
  issn = {0034-6861, 1539-0756},
  doi = {10.1103/RevModPhys.85.961},
  urldate = {2026-08-05},
  copyright = {http://link.aps.org/licenses/aps-default-license},
  langid = {english}
}

@FOOTNOTE{Note1,key="Note1",note="Code and results are publicly available online here:~\protect \href  {https://github.com/olsenlab-science/group1-HFS}{https://github.com/olsenlab-science/group1-HFS}\label {ref:github}"}

\end{document}